# Photoionization and vacancy decay of endohedral atoms

**M. Ya. Amusia**[1, 2]

[1]Racah Institute of Physics, The Hebrew University, Jerusalem 91904, Israel
[2]A. F. Ioffe Physical-Technical Institute, St.-Petersburg 194021, Russia

## Abstract

We demonstrate the role played by the fullerenes shell in the photoionization and vacancy decay of endohedral atoms $A@C_{60}$. It is shown, partly in the frame of a rather simple model that describes the fullerene shell, partly using only the assumption on the smallness of the atom A in comparison to the size of $C_{60}$ that it affects photoionization and the vacancy decay of $A@C_{60}$ profoundly. Namely, it leads to a number of new resonances in photoionization cross-section and other photoionization characteristics as well as strong modifications of the vacancy decay probabilities and to opening of new decay channels.

We will discuss the problem of photon propagation through the $C_{60}$ shell and conclude that at any considered frequency $\omega$, $0 \le \omega \le 60eV$ the $C_{60}$ enhances the incoming radiation. This shows non-metallic dielectric behavior of the 240 collectivized electrons in $C_{60}$.

We discuss also the $C_{60}$ effects upon the fast electron inelastic scattering.

The results obtained are valid qualitatively also for other than $C_{60}$, e.g. for $C_{70}$ or $C_{76}$.

## 1. Introductory remarks

During last decade new atom-like objects consisting of many carbon atoms forming a closed almost "empty" and two-dimensional surface, so-called fullerenes, were discovered. A lot of attention was given to investigation of their structure using photoabsorption data. It appeared that all but the $1s^2$ electrons of carbon atoms are to large extent collectivized. As a result, the photoabsorption spectrum of these objects, just as of atomic nuclei and some multi-electron atoms is characterized by a prominent maximum called *Giant resonance* (see [1, 2] and references therein). The integrated power of this resonance, its total *oscillator strength*, is close to the total number of electrons (or nucleons) in the system under consideration. Although fullerenes are tough objects for *ab initio* theoretical calculations, a number of theoretical approaches were developed to investigate them. Inevitably, they are based on different approximating assumptions.

It appeared that fullerenes have a rather specific and unusual feature: almost any atom from the periodic table can be located inside fullerene, thus leading to so-called *endohedral atoms*. In fact, not only one, but also several atoms can be located inside a fullerene, particularly inside a structure that consists of sixty or even more carbon atoms. These objects are in fact micro getero-structures, and as such deserve particular attention. Indeed, the properties of the inner atom, for instance its photoionization cross-sections and created in such process vacancy decay, is strongly affected by the surrounding fullerene shell. The opposite is also valid and the fullerene is affected by attraction from the atom located inside.

The aim of this paper is to demonstrate how the photoionization cross section and the probability of the subsequent vacancy decay of most studied endohedral atoms $A@C_{60}$ is altered due to the action of the fullerene $C_{60}$. The results will be obtained either using a very simple model, or on the base of existing experimental data.

In this paper we will present two types of effects upon the photoionization cross-section of the atom located inside the fullerene shell: the reflection of the photoelectron wave by the static field of fullerene shell, and incoming photon modification due to dynamic response of the collectivized electrons of the fullerene. Some attention will be given to investigation of conducting properties of $C_{60}$. Then we will discuss the modification by $C_{60}$ shell of the decay of the vacancies created in photoionization of the atom A. We will also investigate the effect of $C_{60}$ shell upon the two-electron photoionization of the He atom, located inside the fullerene shell. The inelastic scattering of the photoelectron with the fullerenes shell will be ignored.

Thus, we will investigate processes that, although in some cases having relatively small cross sections and probabilities, are distinguishable from other in electron-spectroscopy or coincidental experiments. The inelastic scattering of photoelectrons from the inner atom A by the fullerene shell and "closing" of the inner atom vacancies by fullerene electron are not considered in this paper

Atomic System of units, with electron mass *m*, charge *e*, and Planck constant $\hbar$ equal to 1, $m = e = \hbar = 1$ is used in this paper.

**2. Confinement resonances**

For concreteness, let us concentrate on only one type of a fullerene, namely $C_{60}$. Assume for simplicity that the atom A is located at the center of almost spherically - symmetric $C_{60}$ shell. Static action of $C_{60}$ shell is schematically presented in Fig. 1. Note that the action of the fullerene shell upon the photoelectron wave leads to interference of direct and reflected waves as it is shown in Fig. 1 (see [3] and references therein).

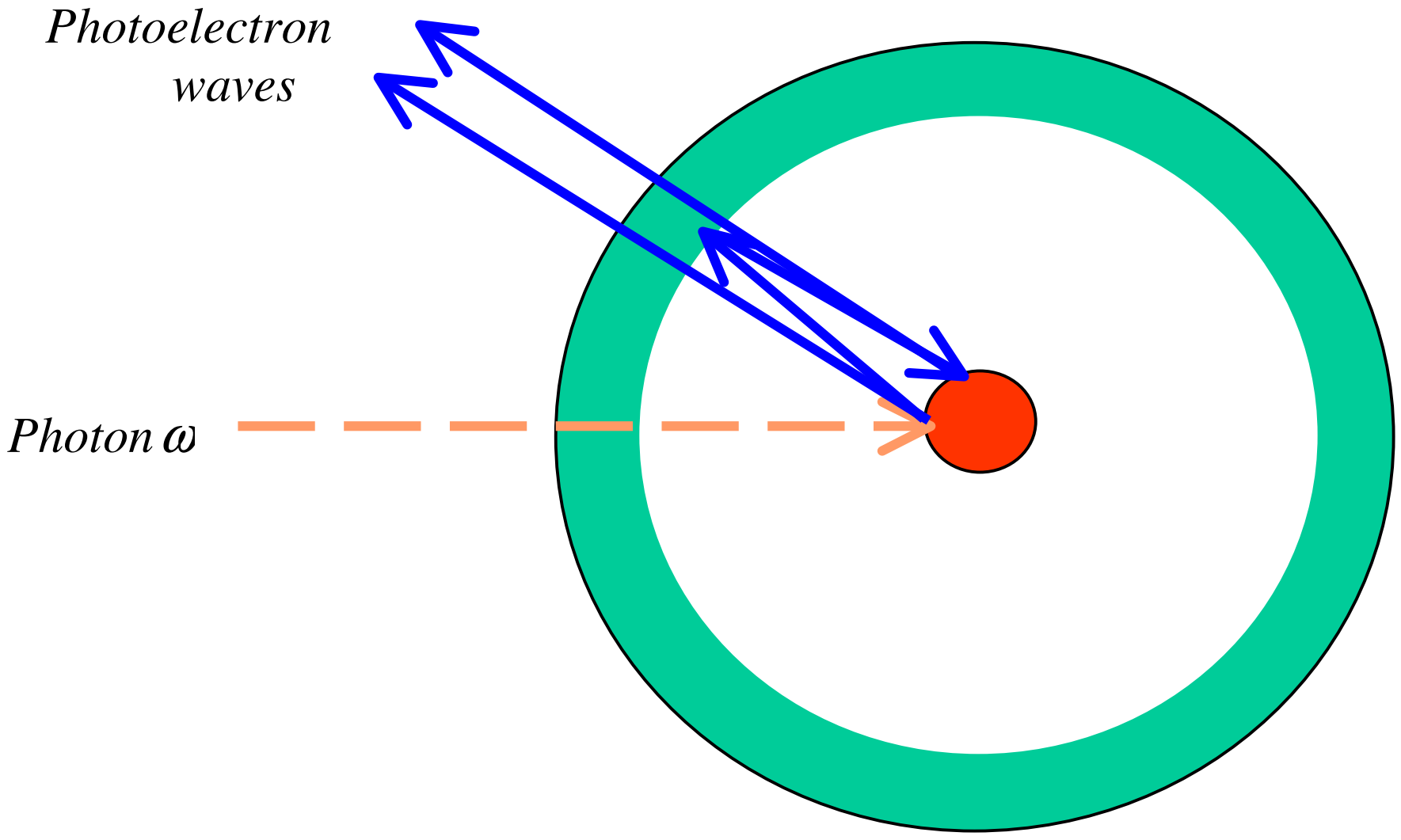

Fig. 1. Static action of $C_{60}$ field upon photoelectron from atom A in $A@C_{60}$

Since the thickness of the $C_{60}$ shell $\Delta$ is much smaller than its radius *R*, for low-energy photoelectrons one can substitute the $C_{60}$ potential by a zero-thickness pseudo-potential [4, 5]:

$$V(r) = -V_0 \delta(r - R). \qquad (1)$$

The parameter $V_0$ is determined by the requirement that the binding energy of the extra electron in the negative ion $C_{60}^-$ is equal to its observable value. The addition of (1) can be taken into account analytically [4, 5], by expressing the partial wave with angular momentum $l$ via the so-called regular $u_{kl}(r)$ and irregular $v_{kl}(r)$ solutions of the atomic Hartree – Fock equation for a photoelectron with linear momentum $k = \sqrt{2\varepsilon}$ , where $\varepsilon$ is the photoelectrons energy, connected to the photon frequency $\omega$ by the relation $\varepsilon = \omega - I_A$ and $I$ is the atom A ionization potential .

Inclusion of (1) leads to a factor $F_{l'}(k)$ in the photoionization amplitude that depends only upon the photoelectron's linear $k$ and angular $l'$ moments [5, 3]:

$$F_l(k) = \cos\delta_l(k)\left[1 - \tan\delta_l(k)\frac{v_{kl}(R)}{u_{kl}(R)}\right], \qquad (2)$$

where $\delta_l(k)$ is the photoelectron's elastic scattering phase shift that can be expressed using the following relation:

$$\tan\delta_l(k) = \frac{u_{kl}^2(R)}{u_{kl}(R)v_{kl}(R) - k/2V_0}. \qquad (3)$$

Using (2), one can obtain the following relation for the partial photoionization cross-sections of the endohedral atom $\sigma_{kl,l'}^{AC}(\omega)$ that corresponds to $l \to l' = l \pm 1$ transitions from the $nl$ subshell:

$$\sigma_{nl,kl'}^{AC}(\omega) = | F_{l'}(\omega) |^2 \ \sigma_{nl,kl'}^{A}(\omega) \qquad (4)$$

where $\sigma_{nl,kl'}^{A}(\omega)$ is the pure atomic cross-section for the same transition. Maxima in $F_l(\omega)$ are called *confinement resonances* [3].

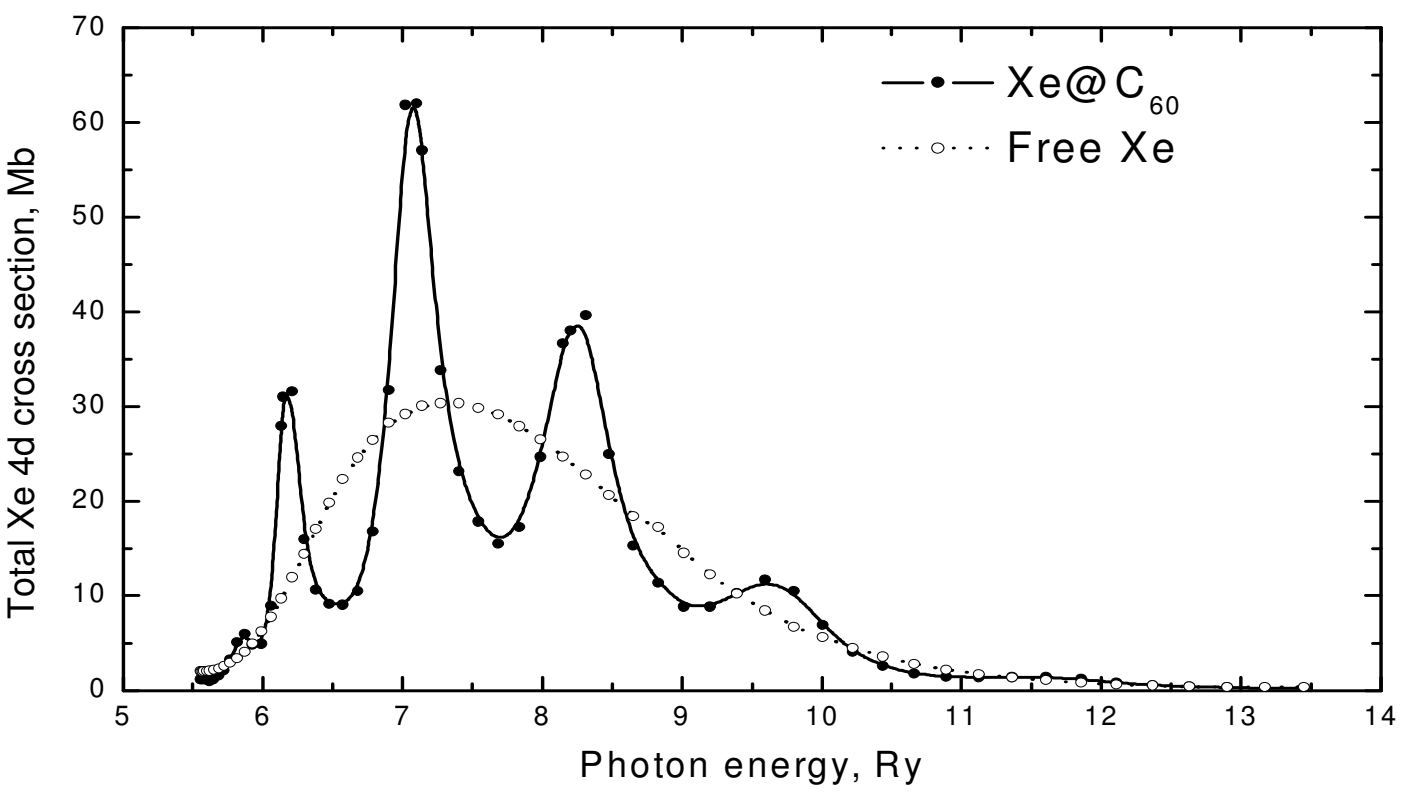

Fig. 2. Destruction of 4d Giant resonance in Xe

We have applied recently this relation to the Giant resonance in the $Xe@C_{60}$ $4d^{10}$ subshell [6, 7]. The results are presented in Fig.2, where it is demonstrated that reflection by $C_{60}$ shell transforms the pure atomic resonance into four relatively narrow ones.

### 3. Resonances in angular distributions

The following relation gives the differential in angle photoionization cross-section of a free atom by non-polarized light [8]:

$$\frac{d\sigma_{nl}(\omega)}{d\Omega} = \frac{\sigma_{nl}(\omega)}{4\pi}[1 - \frac{\beta_{nl}(\omega)}{2}P_2(\cos\theta) + \kappa\gamma_{nl}(\omega)P_1(\cos\theta) + \kappa\eta_{nl}(\omega)P_3(\cos\theta)], \quad (5)$$

where $\kappa = \omega/c$, $P_{1,2,3}(\cos\theta)$ are the respective $l = 1,2,3$ Legendre polynomials, $\theta$ is the angle between photon and photoelectron linear moments, $\beta_{nl}(\omega)$ is the dipole, while $\gamma_{nl}(\omega)$ and $\eta_{nl}(\omega)$ are so-called non-dipole angular anisotropy parameters.

The reflection of photoelectron wave by the fullerene shell affects not only the partial and total cross-sections, but also dipole and non-dipole angular anisotropy parameters. The latter can be expressed via matrix elements, dipole $\tilde{D}_{l\to l\pm1}(\omega) \equiv D_{l\pm1}(\omega)\exp[i\Delta_{l\pm1}(\varepsilon)]$ and quadrupole $\tilde{Q}_{l\to l,l\pm2}(\omega) \equiv Q_{l,l\pm2}(\omega)\exp[i\Delta_{l,l\pm2}(\varepsilon)]$, where $D_{l\pm1}(\omega)$, $Q_{l,l\pm2}(\omega)$, $\Delta_{l\pm1}$ and $\Delta_{l,l\pm2}$ are absolute values of the amplitudes and phases for respective transitions. In general, the corresponding expressions are rather complicated [8]. Since our aim is to illustrate the effect of the fullerene shell, we give here only the expression for atomic *d* subshell and omit for simplicity of notations the arguments $k$ for $F_l(k)$, $\omega$ for $D_l(\omega)$ and $\varepsilon$ for $\Delta_l(\varepsilon)$. Just as in the previous Section we consider the 4*d*-subshell. Then the expression for the dipole angular anisotropy parameter $\beta_l(\omega)$ takes the form [8]:

$$\beta_{4d} = \frac{2}{5[2F_1^2D_1^2 + 3F_3^2D_3^2]}[F_1^2D_1^2 + 6F_3^2D_3^2 - 18F_1F_3D_1D_3\cos(\Delta_3 + \delta_3 - \Delta_1 - \delta_3)]. \quad (6)$$

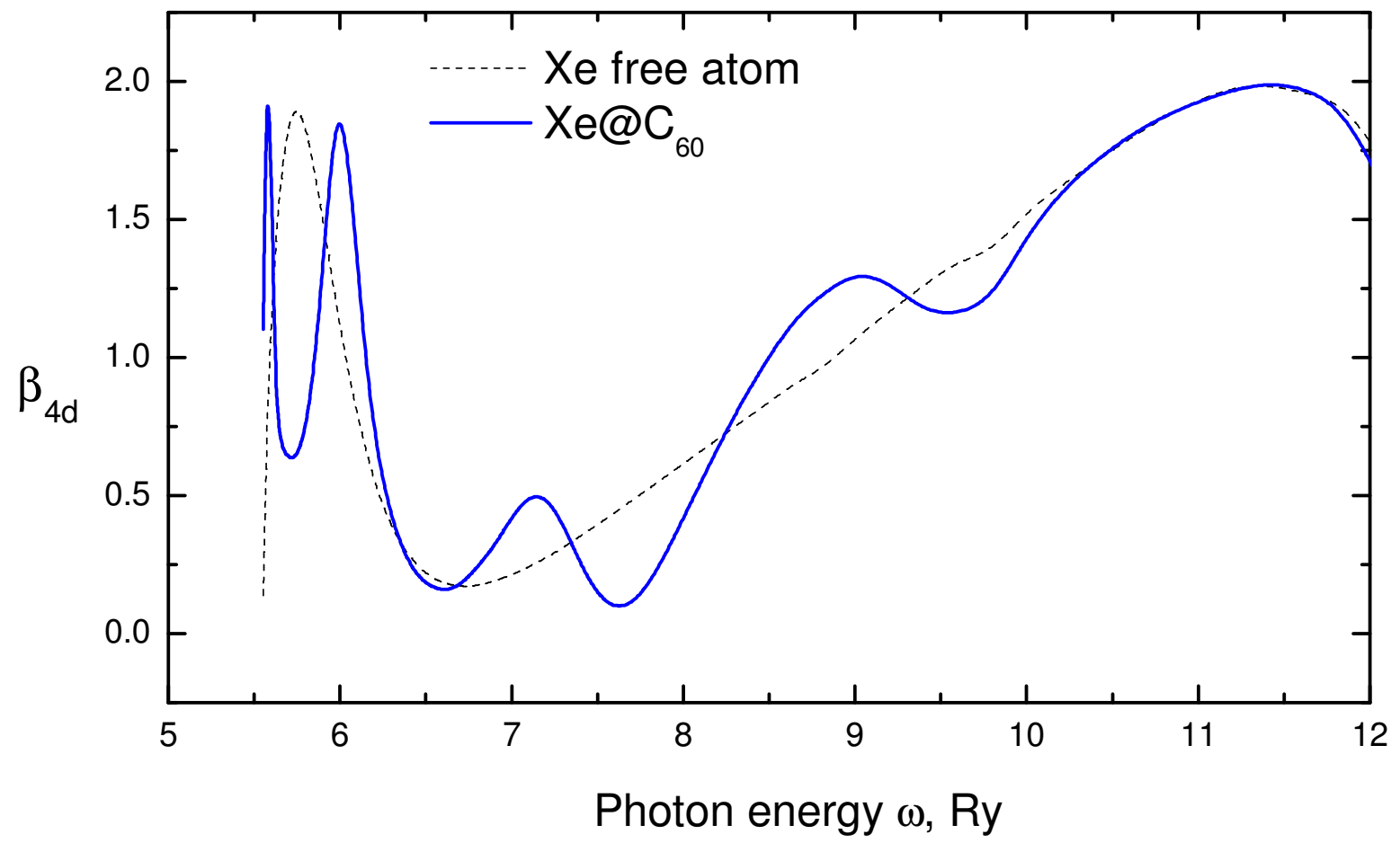

Fig. 3.Dipole Angular anisotropy parameter

It is seen from Fig. 3 that confinement resonances prominently modifies the dipole angular anisotropy parameter.

Non-dipole parameters are represented by expressions considerably more complex than (6). This is why we illustrate their structure by the example of the *s*-shell parameter:

$$\gamma_s = 6\frac{\omega}{c}\frac{F_2}{F_1}\frac{Q_2}{D_1}\cos(\Delta_2 + \delta_2 - \Delta_1 - \delta_1)\,. \tag{7}$$

General formula for any $l$ can be found in [9, 10]. Fig. 4 illustrates the effect of $C_{60}$ shell upon the $\gamma_{4d}(\omega)$ parameter:

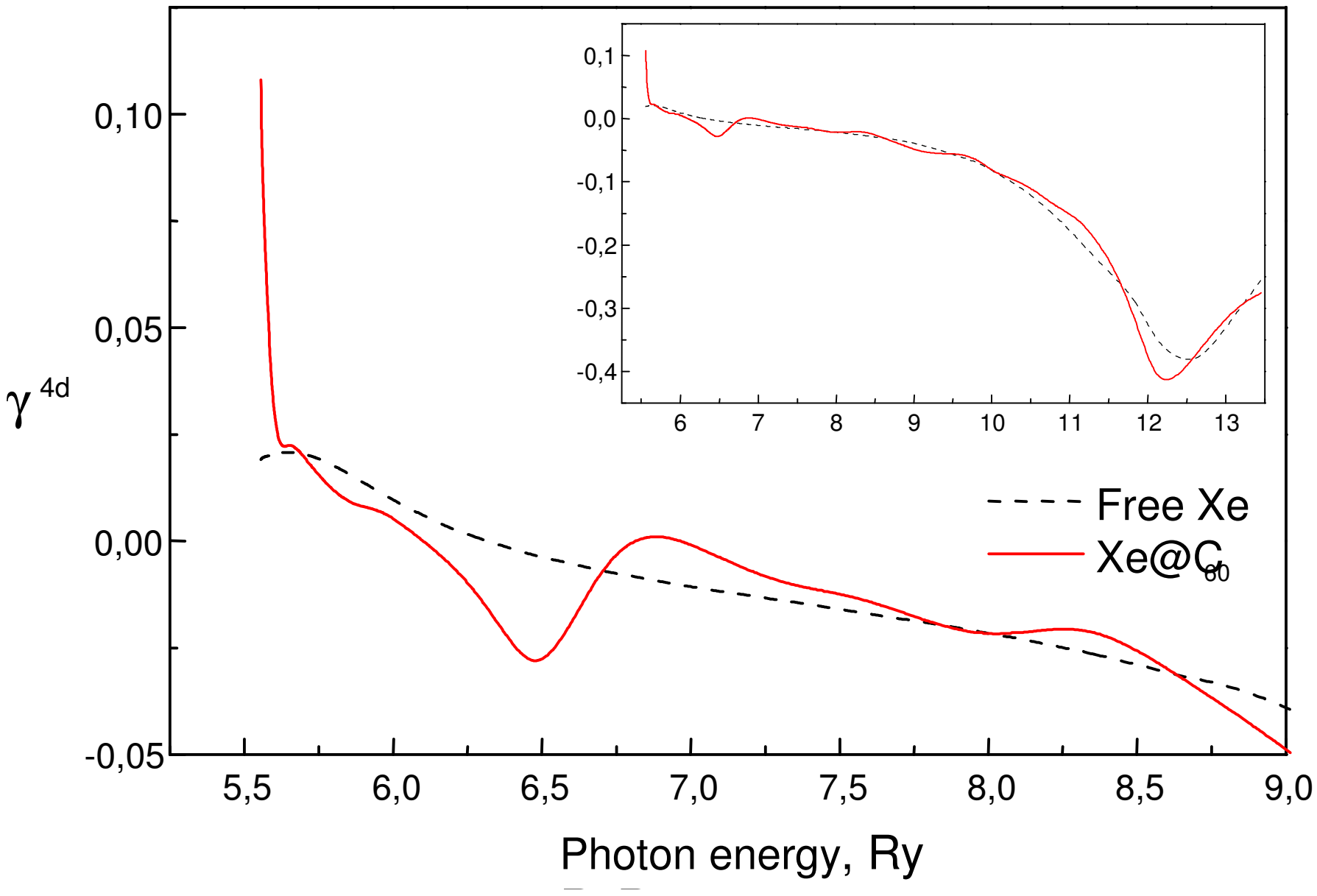

Fig. 4. Nondipole angular anisotropy parameter

**IV. Intershell interaction and interference resonances**

In this section we will concentrate on the dynamic reaction of $C_{60}$ shell upon the photoionization of the inner atom A. In the derivation we will follow the analogy between an ordinary atomic multi-electron subshell and shell of 240 collectivized electrons of $C_{60}$.

It has been already more than thirty years since it was predicted that intra- and inter-shell interactions in atoms could modify considerably the partial photoabsorption cross-sections. It was demonstrated that the so-called intershell interaction leads to *interference resonances* [11, 12].

From physical point of view the situation remains the same, when along with atomic shells the additional fullerene shell appears. The essential simplification is that the interaction between the fullerene and atomic A electrons is sufficient to take into account in the first order of perturbation theory, since the radius of the fullerene $R$ is much bigger than the atomic radius $r_A$. As a result, one has for the main term of this interaction only a dipole-dipole contribution:

$$V_{AC} \approx \mathbf{r}_A \cdot \mathbf{r}_C / R^3, \qquad (8)$$

Here $\mathbf{r}_A$ and $\mathbf{r}_C$ are the radius-vectors of atomic A and fullerene electrons, respectively. This relation is parametrically justified for atoms, which ionization potential is much bigger than that of a fullerene.

In the language of the diagrammatic theory of the many-body systems the photoabsorption amplitude is given by the following equation [11, 12]:

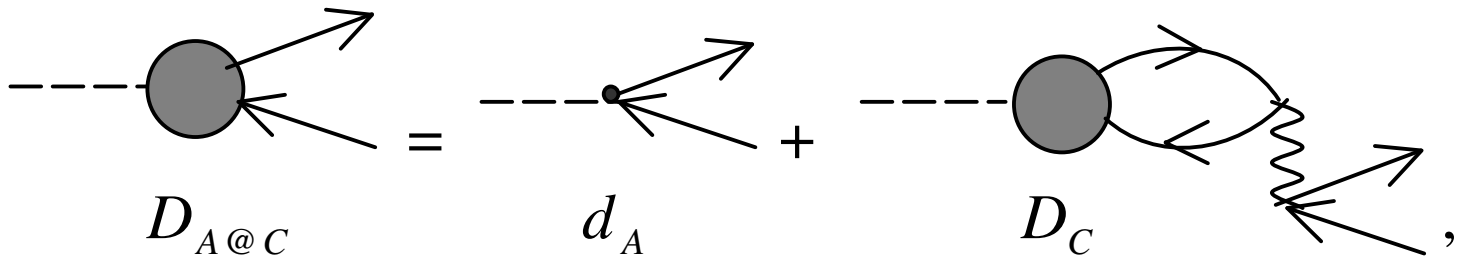

(9)

where $D_{A@C}$, $d_A$, $D_C$ are photoionization amplitudes of $A@C_{60}$, the isolated atom and $C_{60}$, respectively. The dashed line, the line with an arrow to the right (left) and the wavy line represent the incoming photon, electron, vacancy and the direct (exchange) Coulomb “atom A- fullerene” (AC) inter-electron interaction. Analytically, one has for $D_{A@C}$:

$$D_{A@C}(\omega) \cong d_A\left(1 - \frac{\alpha_c(\omega)}{R^3}\right), \qquad (10)$$

where $\alpha_c(\omega)$ is the dipole polarizability of the fullerene shell. It is implied that in (10) $d_A$ includes all essential intra-atomic electron correlations.

The combined action of the static fullerene potential, described in Section 2 and the polarization of $C_{60}$ shell is schematically represented in Fig. 5, where we depict the modification of the incoming photon acting upon the atom A modified due to $C_{60}$ polarization, and the reflection of photoelectron waves, as in Fig. 1:

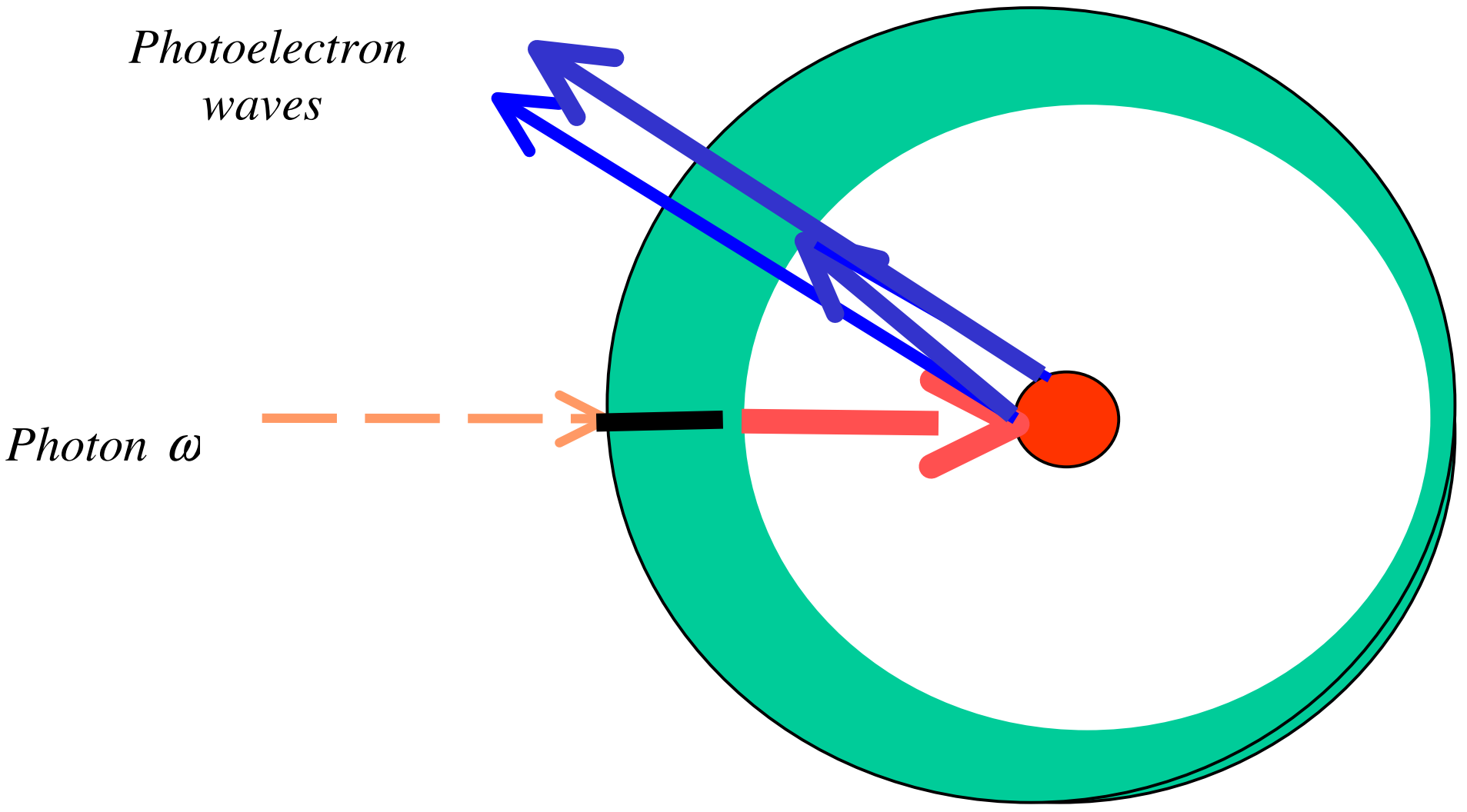

Fig. 5. Schematic representation of the dynamic effect of $C_{60}$ upon $A@C_{60}$ photoionization

Analytically, the partial photoionization cross-section $\sigma^{AC}_{nl,kl'}(\omega)$ of $A@C_{60}$ differs from that of isolated atom A $\sigma^{A}_{nl,kl'}(\omega)$ by two independent factors: $F_l(\omega)$ given by (2) and $G_c(\omega) \equiv \left|1-\alpha_c(\omega)/R^3\right|^2$, thus leading to the following relation:

$$\sigma^{AC}_{nl,kl'}(\omega) = |F_{l'}(\omega)|^2 \left|1-\frac{\alpha_c(\omega)}{R^3}\right|^2 \sigma^{A}_{nl,kl'}(\omega) \equiv |F_{l'}(\omega)|^2 \, G_c(\omega)\sigma^{A}_{nl,kl'}(\omega) . \qquad (11)$$

The expression (10) can be used with experimental cross-section $\sigma^{A}_{kl,l'}(\omega)$ and fullerenes $C_{60}$ polarizability $\alpha_c(\omega)$ and with $F_{l'}(\omega)$, calculated in the frame of better than used above bubble potential (1) model. Note that in the frame of classical approach the introduced above expression for $G_c(\omega)$ was obtained in [13].

## 5. Screening factor

To illustrate the effect of $G_c(\omega)$, let us consider the results of calculations of $\sigma^{AC}_{k0,1}(\omega)$ for the $5s^2$- electrons of Xe in $Xe@C_{60}$ [14]. The polarizability is taken from [15], where it was obtained using the simple, Lorentz-type shape for the $C_{60}$ photoabsorption cross-section. Essential details on the connection of $\alpha_c(\omega)$ and $\sigma_c(\omega)$ are given in Section 7. The factor $G_c(\omega)$ is presented in Fig. 6:

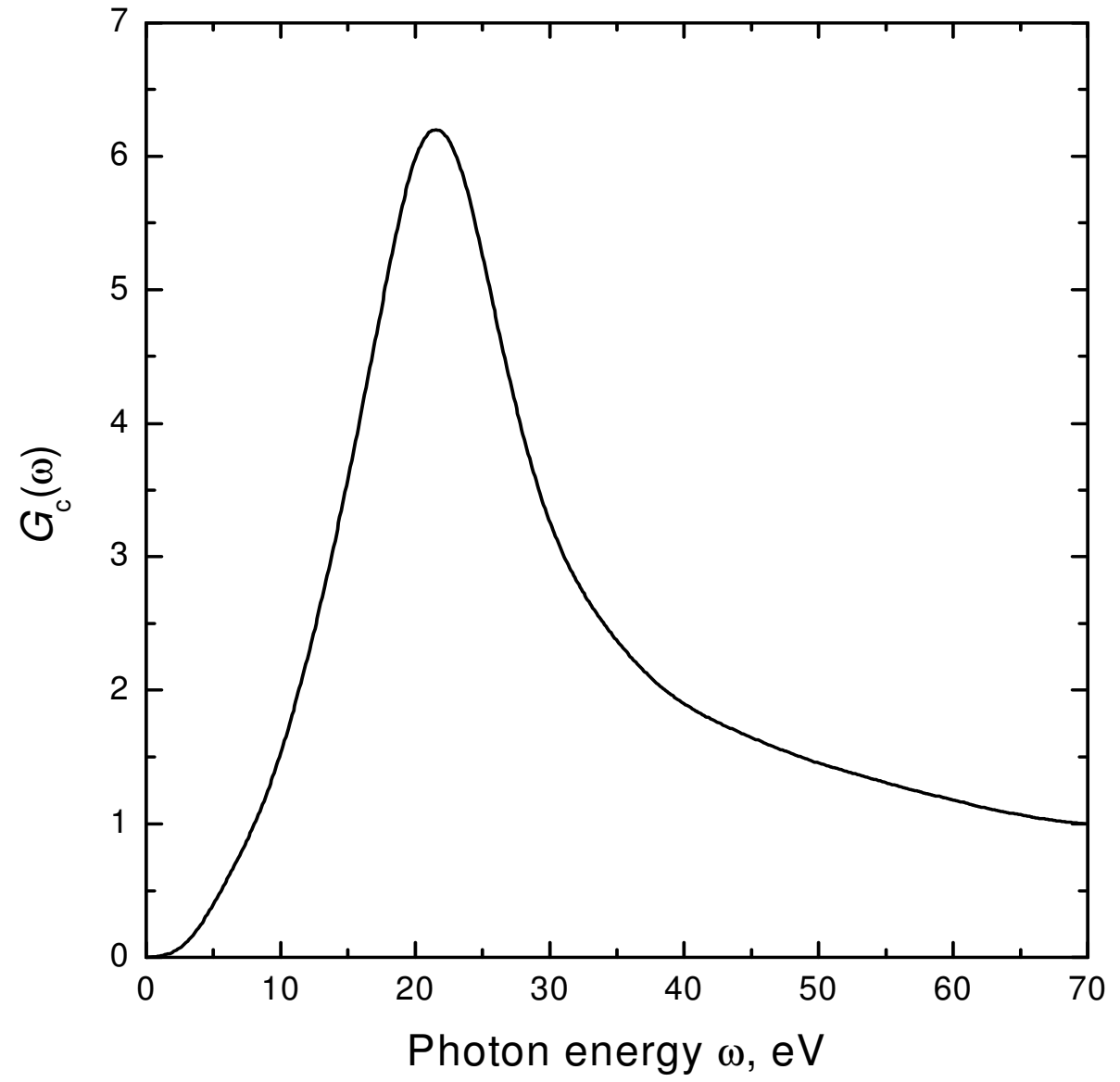

Fig. 6. Factor $G_C$ as a function of photon energy w

It is seen that at low $\omega$ the fullerene is almost completely suppresses the incident electromagnetic field due to polarization of $C_{60}$. As a result, the photoionization cross-section of the encapsulated atom A becomes considerably suppressed. As is seen, the situation is quite different when the photon energy is close to characteristic excitation energy of electrons of the fullerene $C_{60}$ itself, mainly to its Giant resonance that is located at 20-22 eV. A remarkable example here is the Xe 5s subshell photoionization in Xe@$C_{60}$. The ionization potential of this subshell is close to energy of the Giant Resonance of the fullerene. As shown in [14], this results in radical modification of the 5s subshell photoionization cross- section near the threshold.

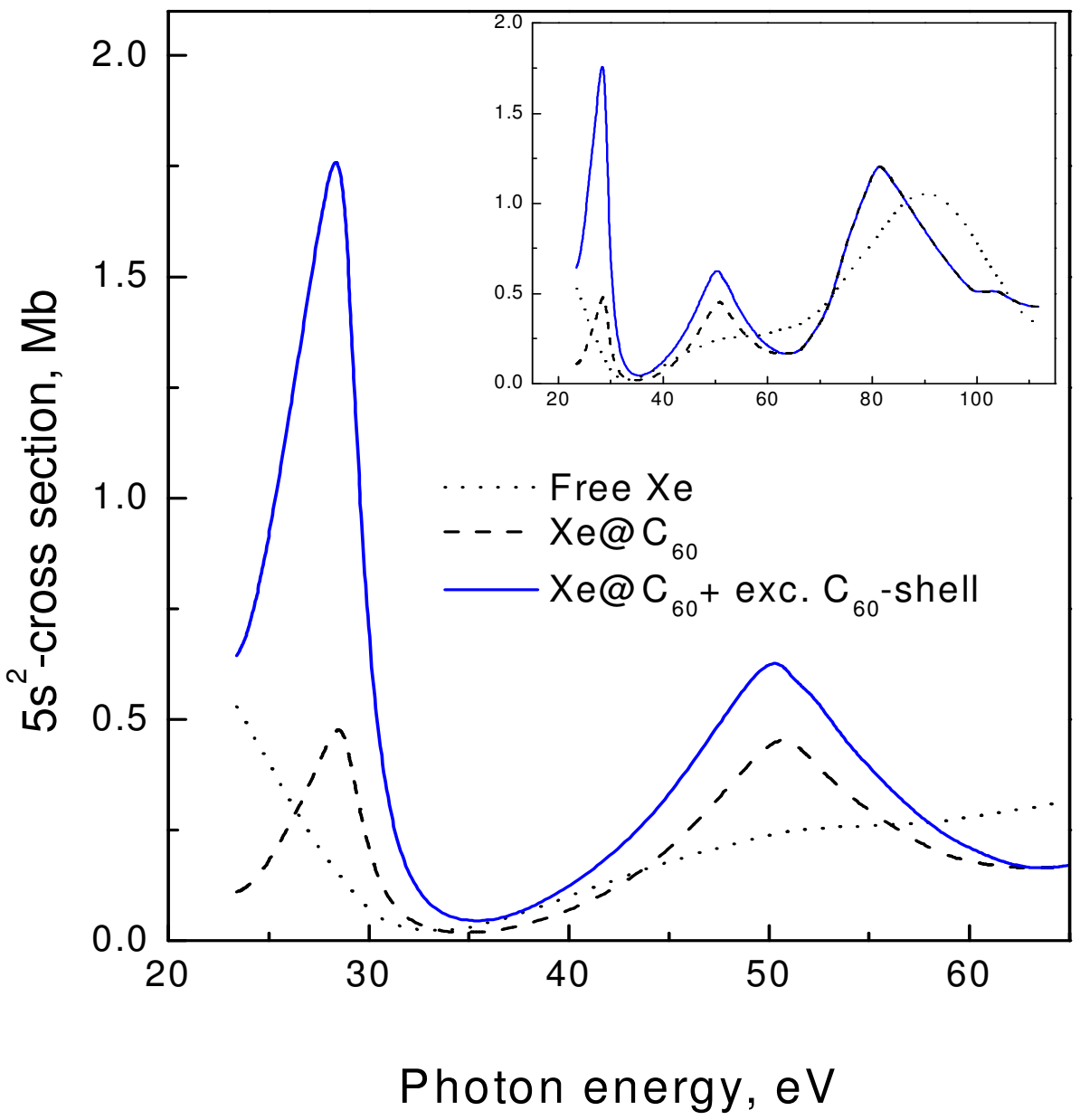

Fig. 7. Photoionization cross-section of 5s-subshell in Xe@$C_{60}$

We see that the 5*s* cross-section with the account of dynamic action of $C_{60}$ (heavy line) is essentially modified as compared to its value that includes the action of $C_{60}$ static potential only (dashed line). The free Xe atom 5*s*-electrons cross-section that is strongly affected by the action of $4d^{10}$ and $5p^6$ subshells is given by the dotted curve [11].

**6. Decay of vacancies**

It is natural to distinguish decays of vacancies in endohedral atoms by the degree of participation of the fullerenes electrons. Namely, the vacancy in the endohedral atom can be occupied by an outer electron either from the same atom or from the collectivized electrons of the fullerene, forming what could be called *atomic* and *fullerene* decays, respectively. If atomic vacancy is deep enough, it can be closed be one of the non-collectivized inner electrons of carbon atoms that form the fullerene, presenting *molecular* decay.

Note, that Auger-decay in collisions of atoms and ions with transition of electrons of one atom into a vacancy in another was suggested about at least forty years ago [16]. Recently intermolecular transitions became an object of extensive consideration (see [17] and references therein).

If the encapsulated atom vacancy decay results in emission of a photon, the fullerene shell can intercept it, then emitting a photon with the same energy. Thus, the vacancy's decay in A occurs through two channels. The first one is the recombination radiation of a photon in the isolated A atom. The second one is the two-step process running through the electron excitation of the fullerene shell. The amplitudes of these channels interfere similar to how this happens in isolated atoms [18] where the decay probability can be altered dramatically from strong enhancement till almost complete suppression. The same situation can happen in the Auger decay: the interaction between electrons, inner and outer, can be modified due to virtual excitations of $C_{60}$ electrons, thus substituting the pure Coulomb inter-electron interaction $V$ by a transferred energy $\omega$ dependent effective one $\Gamma(\omega)$ (see e.g. [12]). The difference between $V$ and $\Gamma(\omega)$ is determined by the so-called intra- or inter-shell effects.

Schematically, these possibilities are presented in Fig. 8. Direct transition of a $C_{60}$ electron into inner vacancy in A is not depicted.

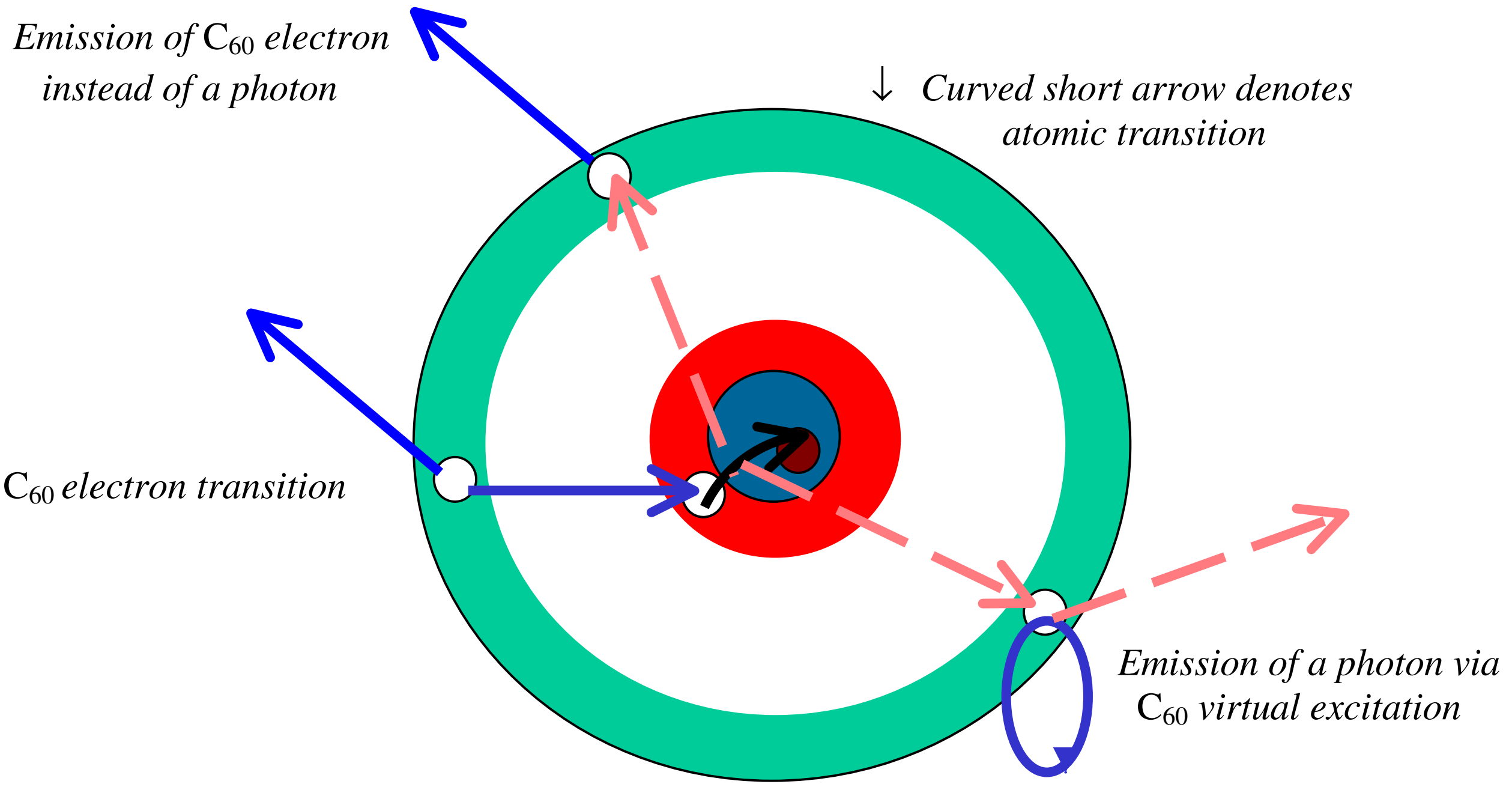

Fig. 8. Schematic representation of $C_{60}$ participation in $A@C_{60}$ vacancy decay

The amplitude of the radiative decay process can be presented by the diagrams [19], similar to the case of radiative decay in isolated atoms [18]:

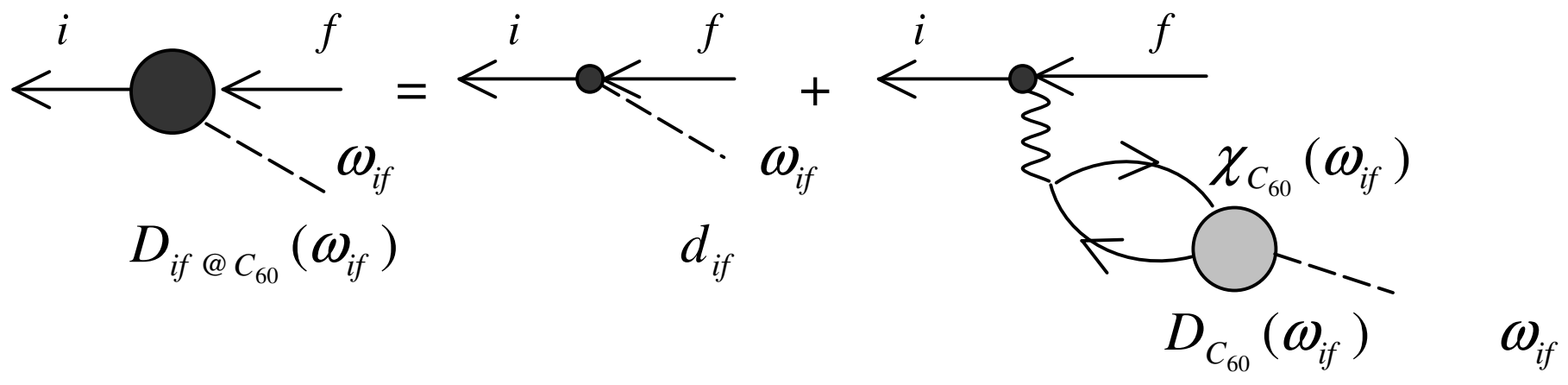

(12)

The essential simplification is the same as in the case of photoionization, considered in Section 4 that is based on approximation (8). This leads to relation (10) for the radiative decay amplitude. As a result, the radiative decay probability for the atomic transition $i \to f$ $\Gamma^{C60}_{\gamma,if}$ is expressed via these same characteristics of the isolated atom $\Gamma^{A}_{\gamma,if}$ and the factor $G(\omega)$

$$\Gamma^{C}_{\gamma,if} = \Gamma^{A}_{\gamma,if}\left|1-\frac{\alpha_c(\omega_{if})}{R^3}\right|^2 \equiv G(\omega)\Gamma^{A}_{\gamma,if}\,. \tag{13}$$

Of special interest is the decay of vacancies in the subvalent *s*- subshells in the endohedral noble gas atoms. These vacancies in isolated atoms are decaying only radiative, while the Auger-decay is energetically forbidden. However, the energy emitted in intra-atomic decay of such a vacancy is sufficient to eliminate one of the collectivized electrons from the $C_{60}$ shell. As a result, the decay probability enhances dramatically, as we will see, by four-six orders of magnitude.

Due to validity of the approximation (8), this Auger-decay can be presented as a photon emission in transition $i \to f$ with the subsequent photoionization by this photon of the $C_{60}$ shell. In the language of diagrams the amplitude of this process can be presented as:

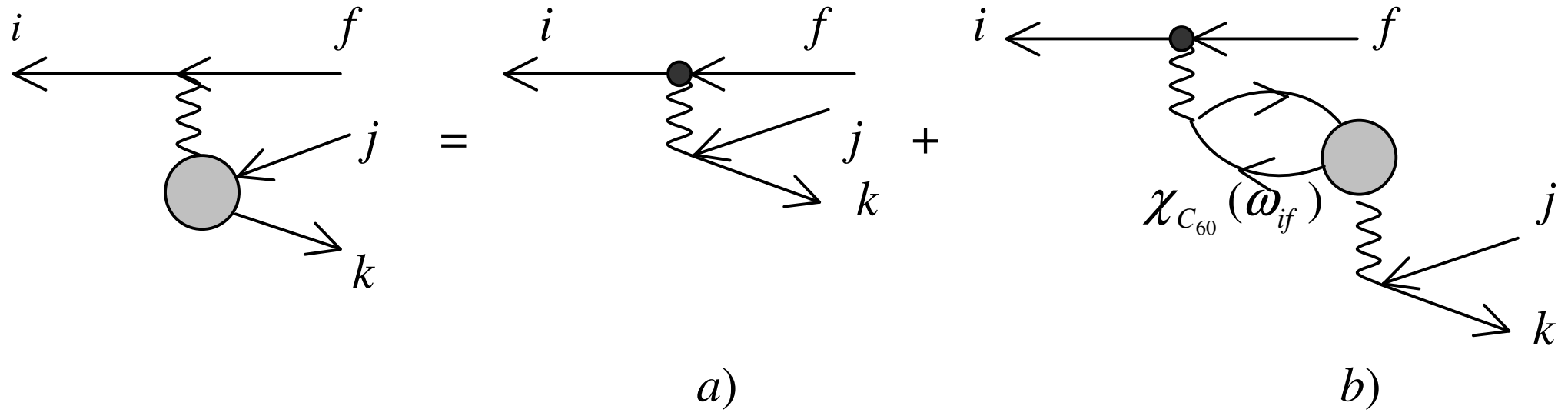

(14)

At first, let us neglect the term *b)* in the right hand side of (14). Then the Auger-decay probability $\Gamma^{C}_{Au,if}$ due to $i \to f$ transition can be expressed via vacancy's atomic radiative width $\Gamma^{A}_{\gamma,if}$ and the photoionization cross-section $\sigma_c(\omega_{if})$ of $C_{60}$ at the transition energy $\omega_{if}$ [19]:

$$\Gamma^{C}_{Au,if} = \Gamma^{A}_{\gamma,if}\,\frac{3}{8\pi}\left(\frac{c}{\omega_{if}}\right)^4 \frac{\sigma_c(\omega_{if})}{R^6} \equiv \Gamma^{A}_{\gamma,if}\,/\,\eta_{AR}(\omega)\,. \tag{15}$$

It is easy to estimate that all but $\Gamma^{A}_{\gamma,if}$ factors are forming a very big quantity, of the order of $10^4 \div 10^6$. Using (13) and (15), one can calculate the so-called *fluorescence yield* $J_R(\omega_{if})$, i.e. the ratio of the radiative to the total decay widths [19]:

$$J_R(\omega_{if}) = \frac{8\pi}{3}\left(\frac{\omega_{if}}{c}\right)^4 \frac{R^6}{\sigma_c(\omega_{if})}\left|1-\frac{\alpha_c(\omega_{if})}{R^3}\right|^2. \tag{16}$$

Of some interest is to estimate whether the virtual excitations of $C_{60}$ electrons can modify the interaction between the atomic electron that undertake transition from the level *f* to *i* and one of the $C_{60}$ electrons. Diagrammatically, this effect is represented by (14b). As it is shown in [19], the respective correction leads to the second term in the square brackets in (17):

$$\Gamma^{C}_{Au,i\to fkj} = \Gamma^{A}_{Au,i\to fkj}\left|1-\frac{1}{2\pi}\sqrt{\frac{3}{2}}\frac{c^2}{\omega_{if}}\frac{\alpha_c(\omega_{if})}{R^6}\left(\frac{\Gamma^{A}_{\gamma,if}}{\Gamma^{A}_{Au,i\to fkj}}\right)^{1/2}[\sigma^{A}_{j}(\varepsilon_k)]^{1/2}\right|^2 . \qquad (17)$$

Estimating this correction term in (17), one can easily show that it is much less than 1 and therefore can be neglected.

**7. Determination of fullerene dipole polarizability**

It is seen that the key quantity that one has to find in order to determine the vacancies decay is the fullerenes polarizability $\alpha_c(\omega)$. It can be calculated *ab initio* in the frame of some models. But we prefer to use the experimental total photoabsorption cross-section $\sigma_c(\omega)$ that is simply proportional to the imaginary part of the $C_{60}$ dipole polarizability $\mathrm{Im}\,\alpha_c(\omega) = c\sigma_c(\omega)/4\pi\omega$. The $\omega$ dependence of the photoabsorption cross-section $\sigma_c(\omega)$ of $C_{60}$ exhibits a powerful maximum at the Giant Resonance energy that is at about 22 eV. Although the experiments [1] provide no direct absolute values of $\sigma_c(\omega)$, it can be reliably estimated by using some different normalization procedures, e.g. that based on the sum rule $(c/2\pi^2)\int\sigma_c(\omega)d\omega = N$ [2]. Here $N$ is the number of collectivized electrons. For $C_{60}$ $N$ was considered to be $N = 240$, i.e. 4 electrons collectivized from each C atom. Using the photoabsorption data from [2] as most reliable, we obtained $N_{eff} \approx 250$.

Note that since the one-electron photoionization cross-section of the positive ion $C^{+}_{60}$ measured and calculated in Ref. [20] is much smaller than the cross section of the neutral fullerene, it cannot be used to define the dynamical dipole polarizability of the neutral $C_{60}$. The dispersion relation

$$\mathrm{Re}\,\alpha_c(\omega) = \frac{c}{2\pi^2}\int_{I_c}^{\infty}\frac{\sigma_c(\omega')d\omega'}{\omega'^2-\omega^2} \qquad (18)$$

connects the real part of the polarizability $\mathrm{Re}\,\alpha_{C_{60}}(\omega)$ with the photoionization cross-section. In Eq. (14) $I_c$ is the fullerenes ionization potential. The calculation results of the real and imaginary parts of the dynamic polarizability of the fullerene shell $C_{60}$ are given in Fig.9 [19]. The experimental photoabsorption cross-section for $C_{60}$ taken from [2] is presented in insertion to Fig. 10. As seen from this figure, the cross-section is small at the threshold (which also means relatively low intensity of discrete excitations) and is dominated by the Giant Resonance well above the threshold. Therefore, in (18) a small contribution of discrete excitations of the $C_{60}$ collectivized electrons are neglected.

The frequency dependence of the imaginary part, as it should be, is similar to the frequency dependence of $\sigma_{C_{60}}(\omega)$. A small peak in the cross-section for photon energy ~7 eV is transformed into a significant maximum, which is explained by the small value of corresponding photon energy as compared to the energy of the Giant Resonance, ~22 eV.

The real part of the polarizability obtained as a result of integration according to formula (16) decreases with the raise of radiation frequency $\omega$, as it should be, as $\mathrm{Re}\,\alpha_{C_{60}}(\omega) \sim -N_{eff}/\omega^2$. The obtained dynamical polarizability is used in calculations of the ratio of "radiative-to-non-radiative" decays' widths $\eta_{RR}(\omega)$ of vacancies in the endohedral atoms that is according to (11) equal to $G(\omega)$, $\eta_{RR}(\omega) = G(\omega)$.

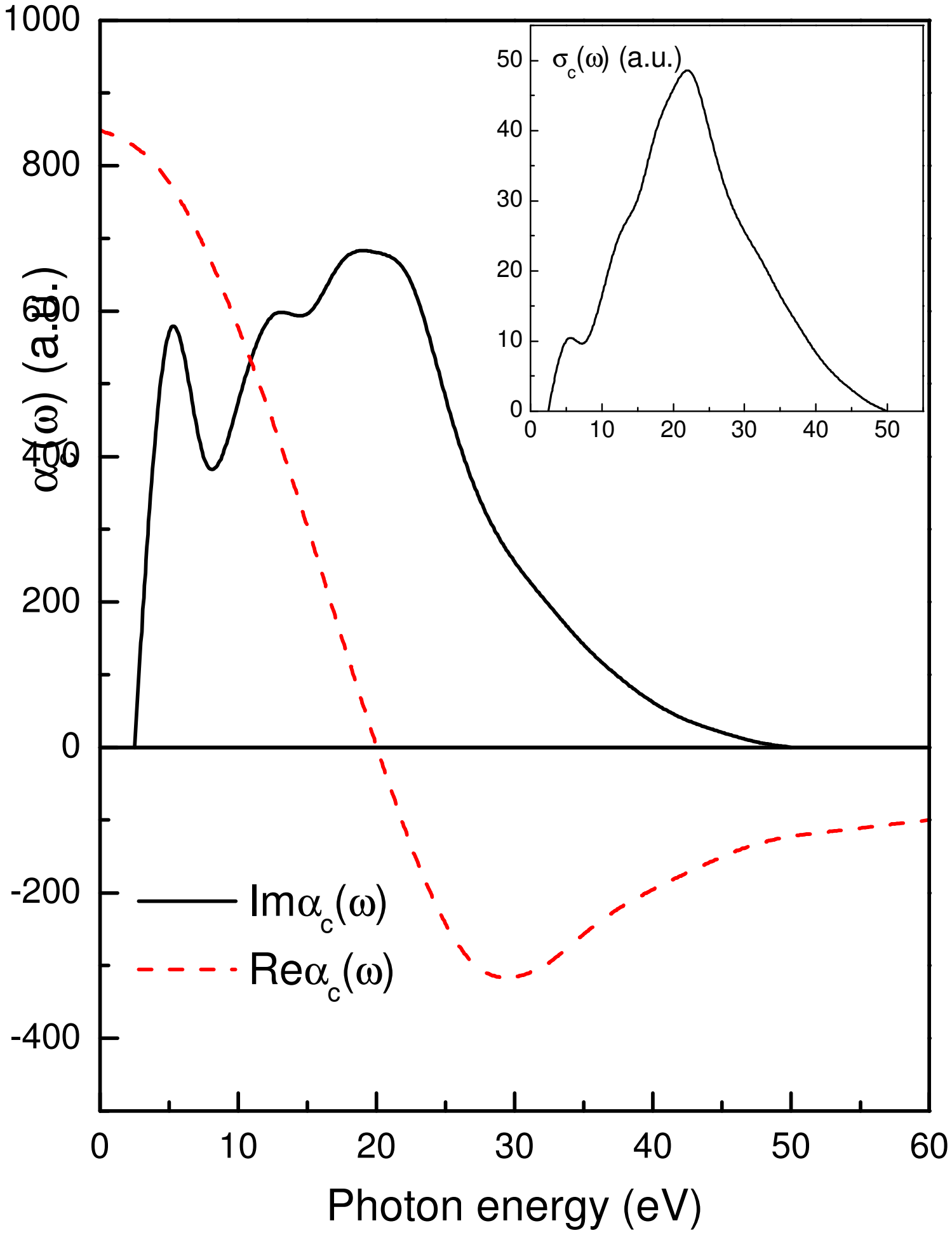

Fig. 9. Real and imaginary parts of dynamical polarizability of $C_{60}$.

The curve $\eta_{RR}(\omega)$ was obtained in [19] and is presented in Fig. 10. Taking the energies for considered in this paper transitions, namely 2*s* – 2*p* in Ne, 3s - 3p in Ar, 4*s* – 4*p* in Kr, and 5*s* – 5*p* in Xe from [21], one obtains the corresponding values for $\eta_{RR}$. They proved to be large enough, varying from 3.5 to 6.2. The $\eta_{RR}$ curve is dominated by two maximums, the bigger due to the Giant resonance at about 22 eV, while the smaller is a consequence of the peculiarity at about 7 eV in the photoionization cross-section. .

In Fig. 10 marks transitions: $\Delta E_1 = 26.85 eV$ corresponds to the 2*s* – 2*p* in Ne; $\Delta E_2 = 13.50$ eV is the 3s - 3p in Ar or 4*s* – 4*p* in Kr; $\Delta E_3 = 11.20$ eV and $\Delta E_4 = 9.90$ eV are the 5*s* – $5p_{3/2}$ and 5*s* – $5p_{1/2}$ transitions in Xe, respectively. All transition energies are taken from [21].

In Fig. 11 we depict the ratio of Auger to the radiative decay widths that according to (15) is given by $\eta_{AR}$ [19]. We see that this enhancement factor increases the decay probability by factors $2\cdot10^4 \div 4\cdot10^5$, bringing the widths to their normal Auger-decay values, which they would have if were not forbidden by energy conservation law. A big maximum is seen in $\eta_{AR}(\omega)$ due to existence of the Giant resonance in the photoionization cross-section of the fullerene itself.

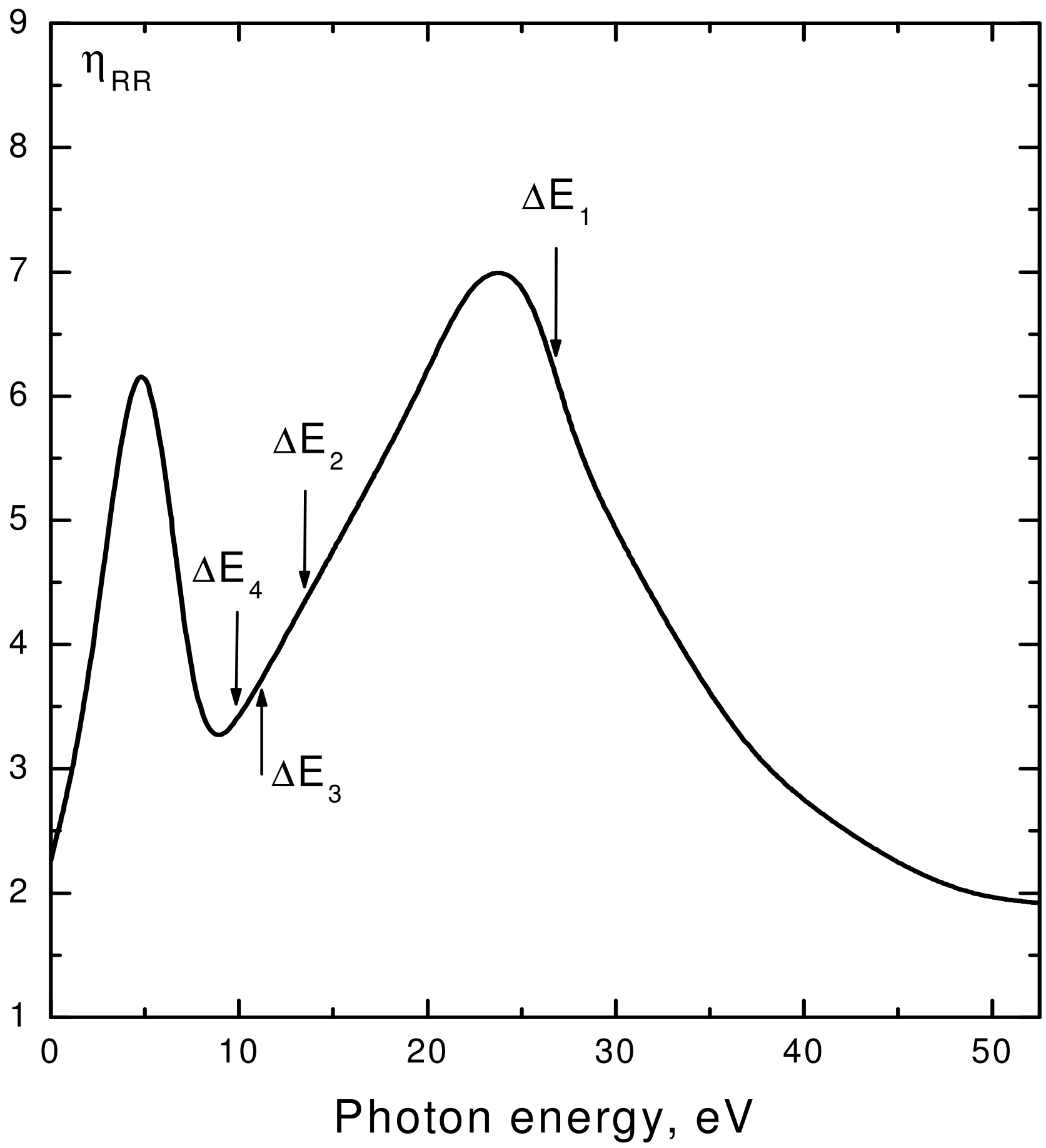

Fig. 10. The ratio of the radiative widths for the same transition in endohedral and free atom.

In Fig. 12 we present the results for the fluorescence yield of the same endohedral atoms and transitions as considered above, namely 2*s* – 2*p* in Ne, 3s - 3p in Ar, 4*s* – 4*p* in Kr, and 5*s* – 5*p* in Xe. This value decreases from Ne, where it is about $4\times10^{-4}$ to Xe, where it is $10^{-5}$ [19]. The meaning of arrows is the same as in Fig. 10.

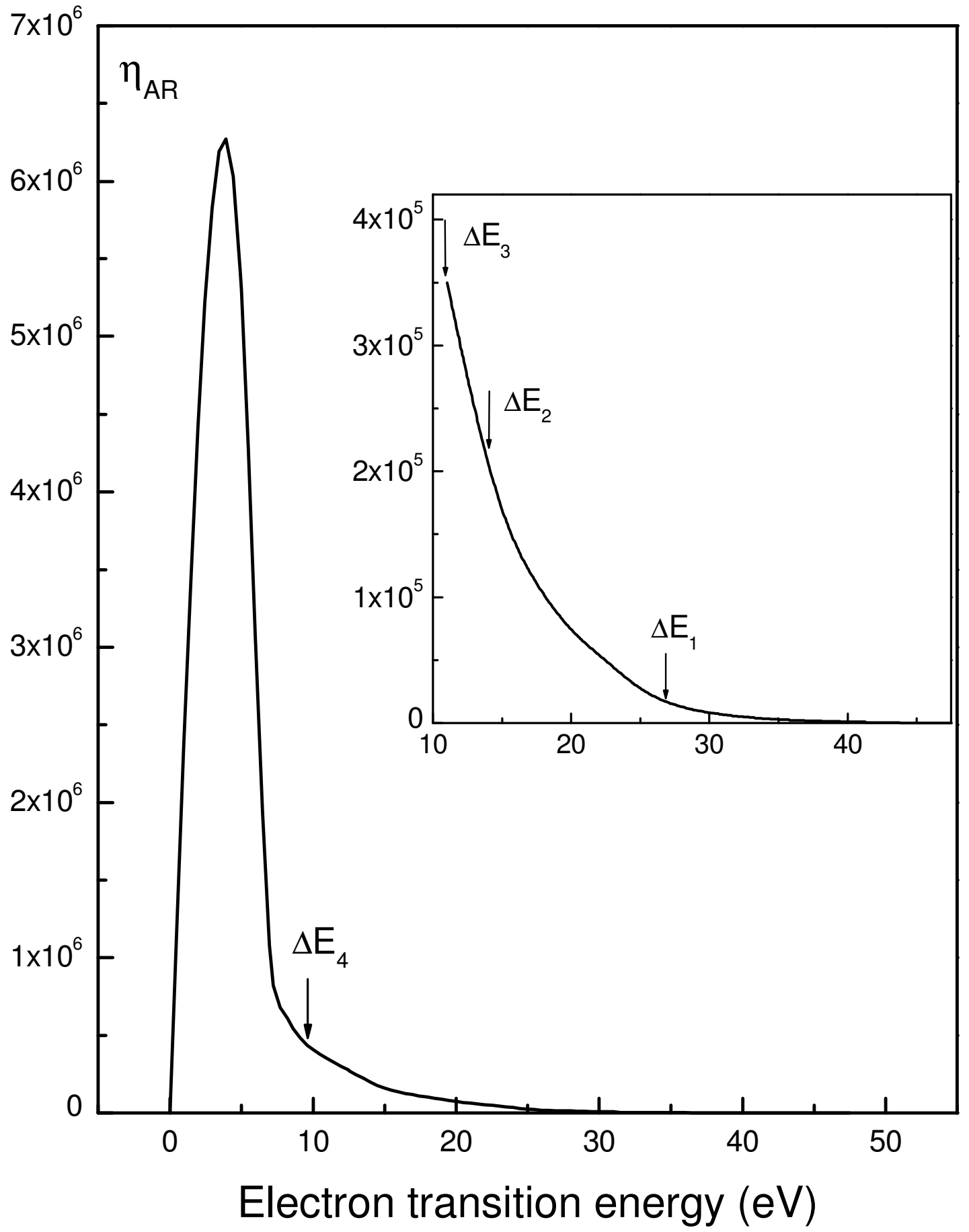

Fig. 11. The ratio of Auger and radiative widths

## 8. Is $C_{60}$ a metal?

Since we have investigated above the reaction of the collectivized electrons of $C_{60}$, we can find out whether they behave as a metal, dielectric or in another way. In fact, we can determine a metal as a system, inside of which the electric field **E** is zero irrespectively to its magnitude outside of the system. So, it is natural to introduce the *field modification* factor [22]

$$\lambda_c(\omega) \equiv | \mathbf{E}_{eff}(\omega) / \mathbf{E}(\omega) |, \tag{19}$$

which for a metal is zero at $\omega = 0$, $\lambda(0) = 0$. Conclusion that $C_{60}$ is a metal has been made in [13]. As it follows from (10), for $C_{60}$ and similar objects, one has

$$\lambda_c(\omega) = \left| 1 - \frac{\alpha_c(\omega)}{R^3} \right|. \tag{20}$$

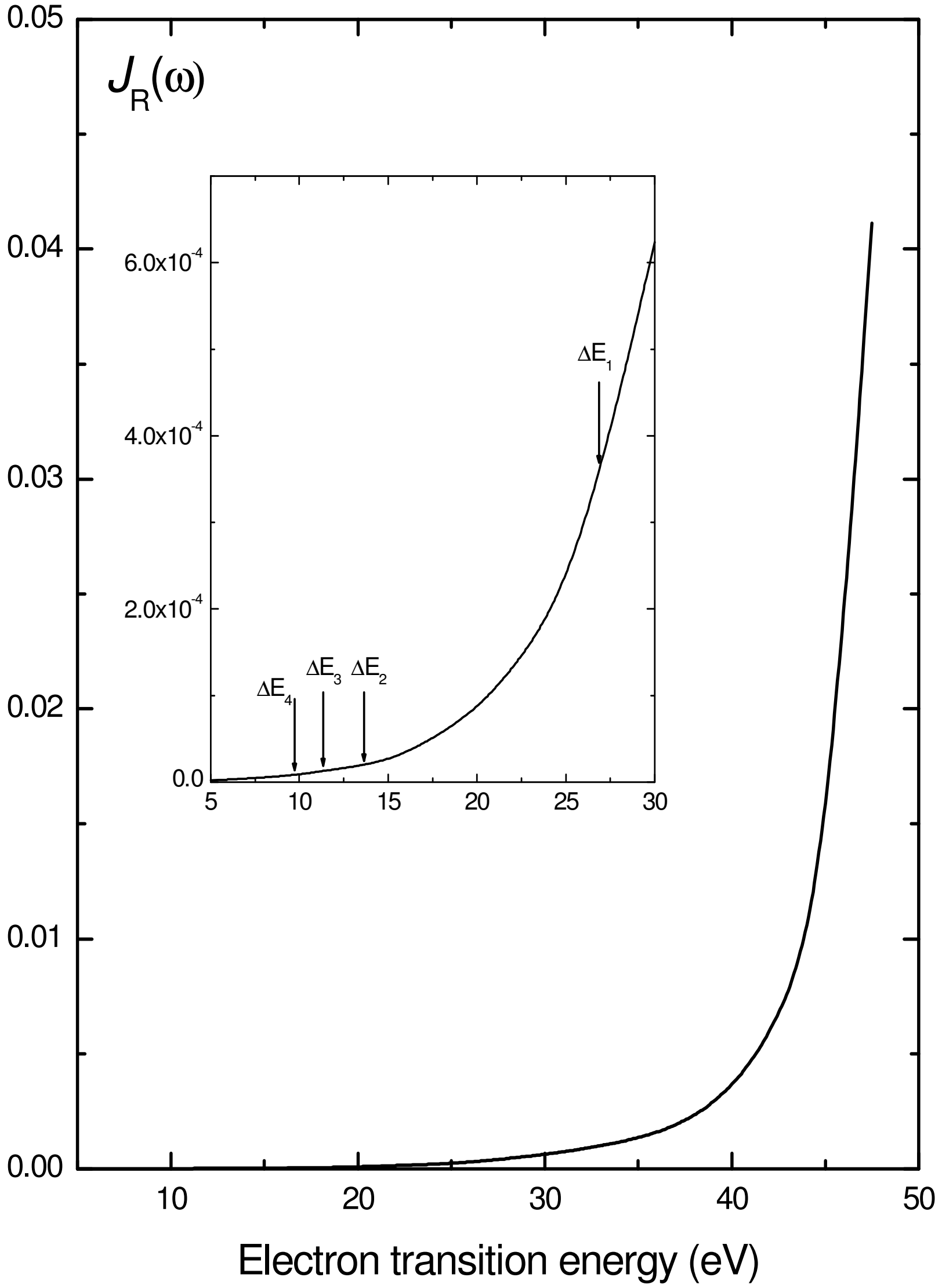

Fig. 12. The fluorescence yield $J_R(\omega)$

According to (11), the cross-section modification factor $G_c(\omega) = \lambda^2(\omega)$ is completely determined by the fullerenes polarizability. It was assumed in [13] that $\mathrm{Im}\,\alpha_c(\omega)$ has a shape of a smooth Lorentz-type profile that satisfies the sum rule [see p. 11], instead of using the best

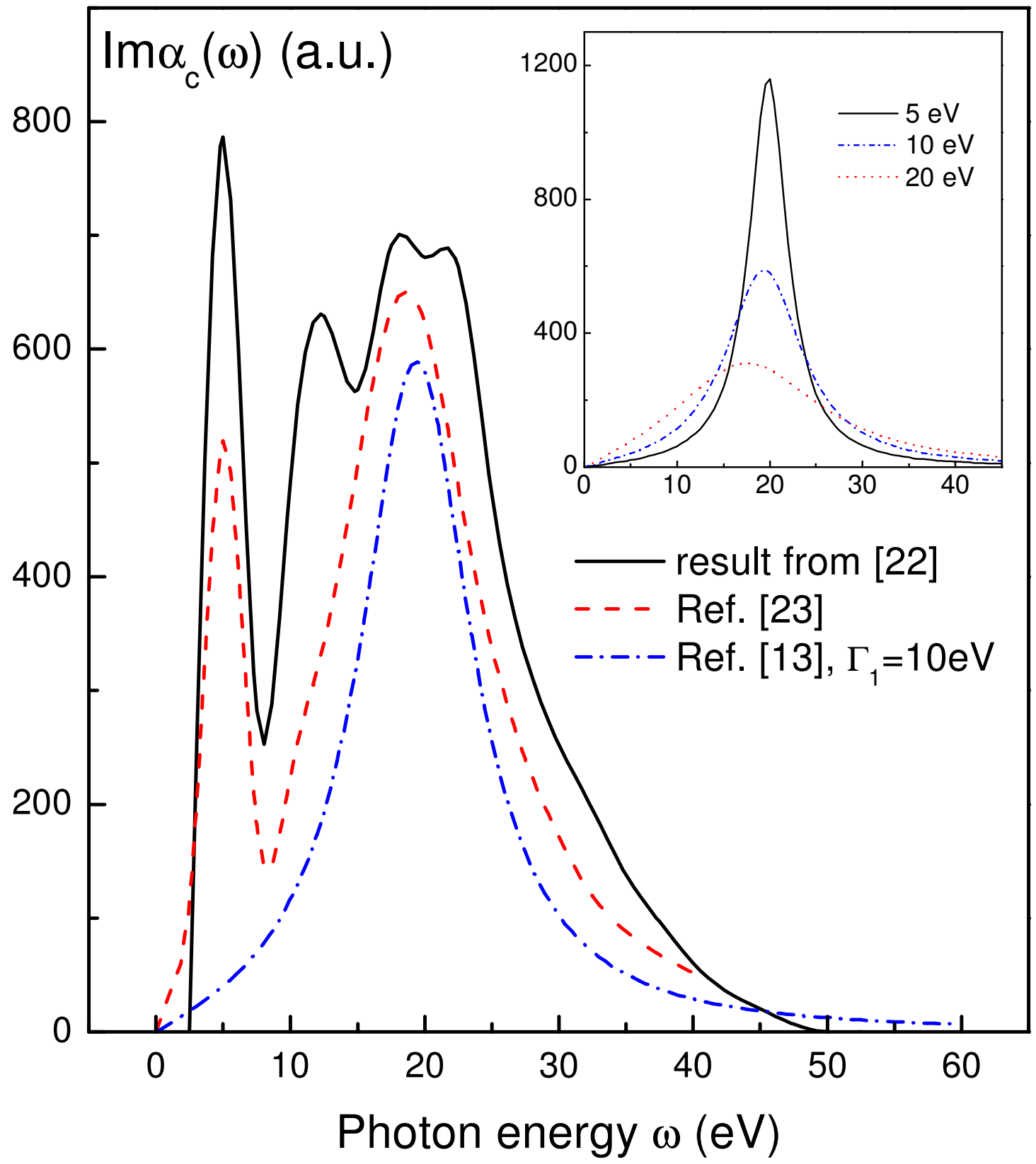

Fig. 13. Imaginary part of the $C_{60}$ polarizability

available $C_{60}$ photoionization cross-section

[2]. In [23] dielectric behavior of free $C_{60}$ electrons was assumed and on this bases the photoionization cross-section of $C_{60}$ was predicted. With these data, we have calculated the imaginary part of the polarizability that is depicted in Fig. 13. In the insertion we present $\mathrm{Im}\,\alpha_c(\omega)$ derived in [13] using a number of widths $\Gamma_1$, considered as a parameter. We see that the difference between $\mathrm{Im}\,\alpha_c$ obtained in the frames of these approaches is big. This is a manifestation of the prominent difference between the experimental cross-sections and the approximate ones that is illustrated by Fig.14.

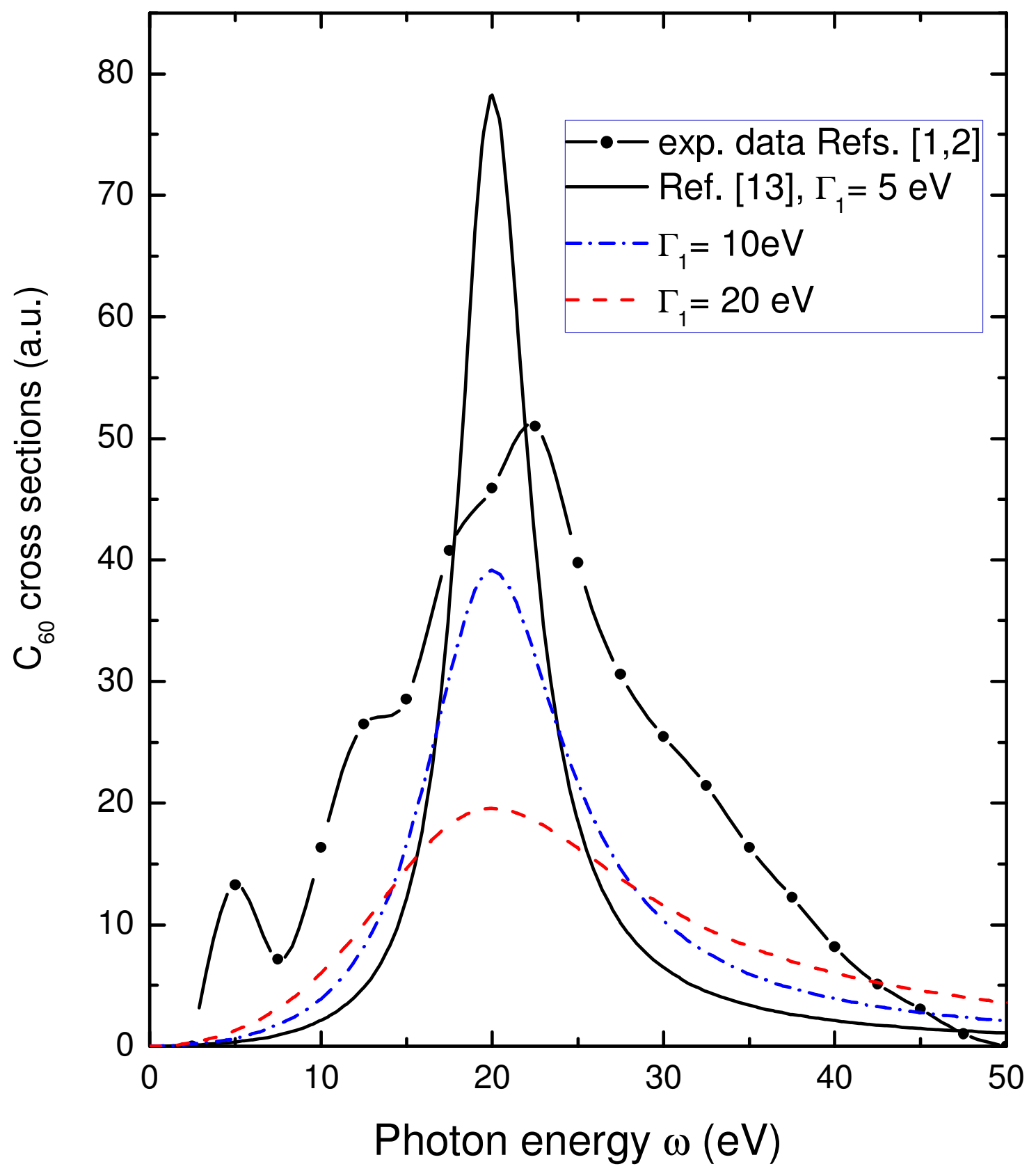

Fig. 14. Photoionization cross-section of $C_{60}$

Using (18), one obtains $\mathrm{Re}\,\alpha_c$ that is presented in Fig. 15. In Fig. 16 we present results for the cross-section modification factor $G_c(\omega)=\lambda^2(\omega)$ and for the field modification factor $\lambda(\omega)$. It is seen that only for a symmetric Lorentz-type profile from [13] one has $\mathrm{Re}\,\alpha_c(0)=R^3$ that leads to metal-type behavior of the $C_{60}$ shell, with strong suppression of the inner field as compared to the external one up to $\omega=10eV$ and its strong enhancement in the region of the Giant resonance maximum. As is seen from the insertion in Fig. 15, the enhancement depends upon the width of the Giant resonance maximum decreasing with its growth.

Calculated with experimental cross-section, the modification factors demonstrate clearly non-metallic behavior at any $\omega$ considerably increasing the field inside $C_{60}$. As it is seen from Fig. 15, the static polarizability of $C_{60}$ is equal to $\alpha_C(0)=849$. Note that this value for a free C atom is equal to $\alpha^c(0)=14$ [24]. Therefore, by the order of magnitude, the polarizability of the 60 carbon atoms is the sum of their polarizabilities: $\alpha^{60C}(0)=14\times 60=840$.

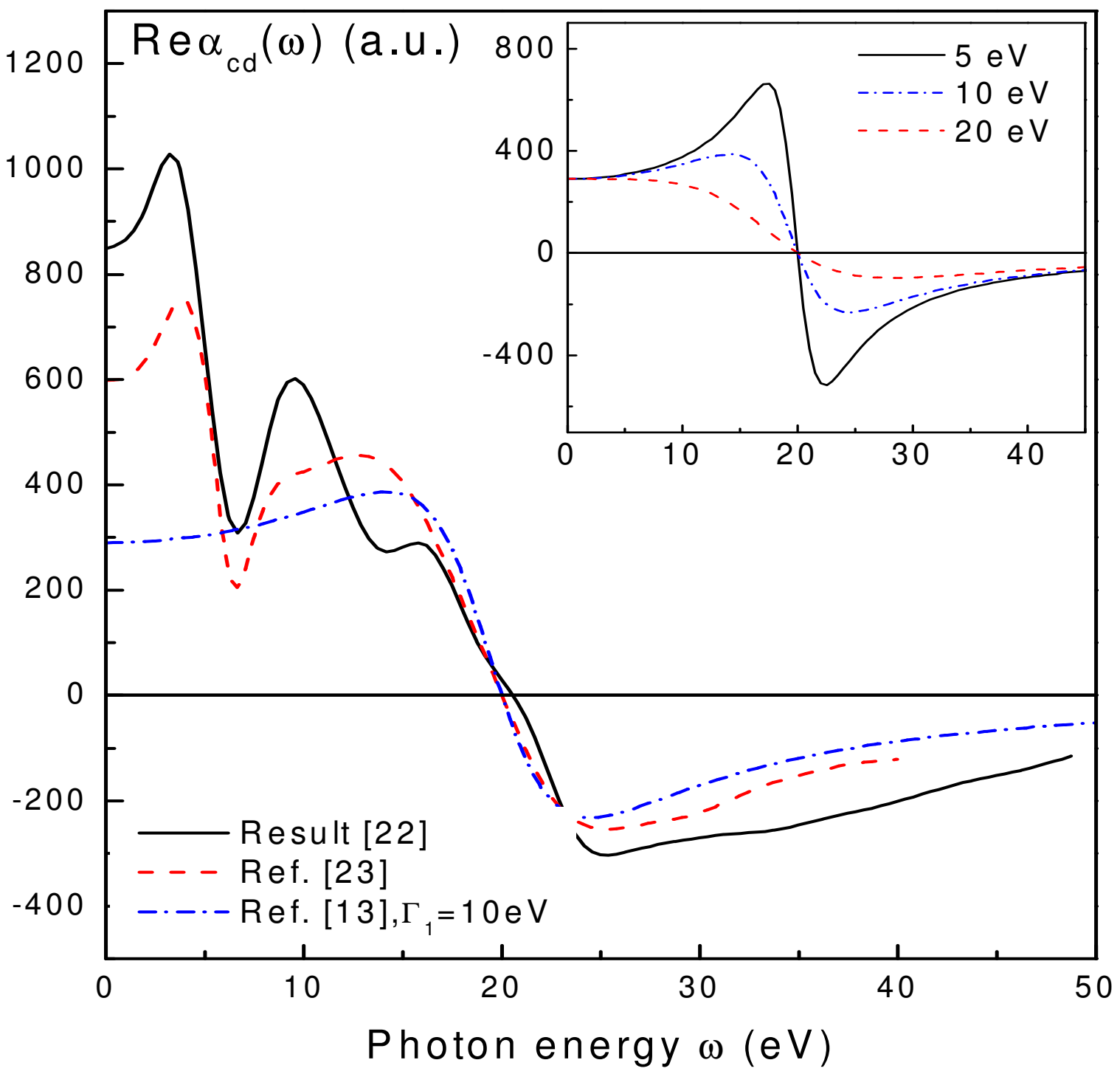

Fig. 15. Real part of $C_{60}$ polarizability

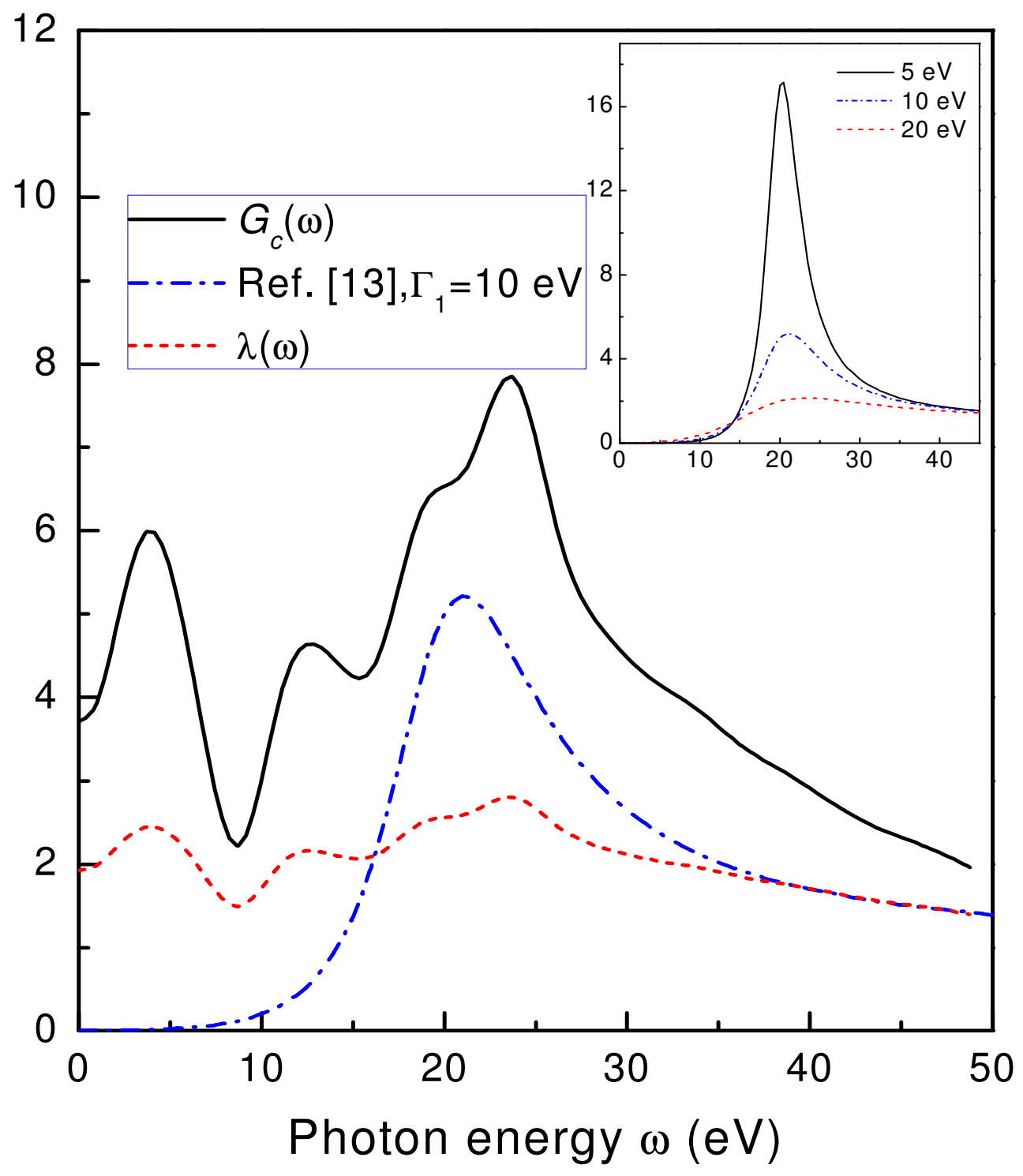

Fig. 16. Modification factors $G_c$ and $\lambda$

**9. Two-electron photoionization**

In this Section we will investigate the two-electron photoionization of endohedral atoms. We consider a process, in which one photon eliminates two electrons just from the inner atom in $A@C_{60}$. As a concrete example we will consider $He@C_{60}$. For this atom the two-electron photoionization threshold $I^{++}$ is high enough, so the deviation of $G_c(\omega)$ from 1 is negligible. Therefore only the reflection of outgoing electron by the potential (1) has to be taken into account. This can be achieved by altering one or two outgoing photoelectron wave functions introducing factors $F_l(k)$ from (2).

If the incoming photon energy is high enough, $\omega >> I^{++}$, the main contribution to the two-electron photoionization cross-section $\sigma_A^{a++}(\omega)$ comes from that region of the photoelectrons' energy spectrum, where one electron is fast and the other slow. The slow electron is represented by an $s$-wave, thus carrying away zero angular momentum. As a result, one has for the endohedral atom two-electron differential in photoelectron energy $\varepsilon$ photoionization cross-section $d\sigma^{++}(\omega)/d\varepsilon$ the following relation:

$$d\sigma^{++}(\omega)/d\varepsilon = |F_o(k)|^2 \, d\sigma^{a++}(\omega)/d\varepsilon\,, \tag{21}$$

where $k = (2\varepsilon)^{1/2}$.

For lower $\omega$, but still well above $I^{++}$, one of the outgoing from $He@C_{60}$ electrons predominantly has angular momentum zero, while the other has one. As a result, instead of (21), it is obtained

$$d\sigma^{++}(\omega)/d\varepsilon = |F_o(k)|^2 |F_1(k')|^2 \, d\sigma^{a++}(\omega)/d\varepsilon\,, \tag{22}$$

where $k' = [2(\omega - I^{++} - \varepsilon)]^{1/2}$.

With further decrease of $\omega$ one has to take into account the exchange between outgoing electrons, which instead of (22) leads to the following relation

$$d\sigma^{++}(\omega)/d\varepsilon = |F_o(k)F_1(k') - F_o(k')F_1(k)|^2 \, d\sigma^{a++}(\omega)/d\varepsilon\,, \tag{23}$$

In Fig. 15 we depict results of calculations for factors $F_o(k)$ and $F_1(k)$. It is seen that they are represented by almost identical oscillating function with the amplitude decreasing while phase shift increasing with the growth of $k$.

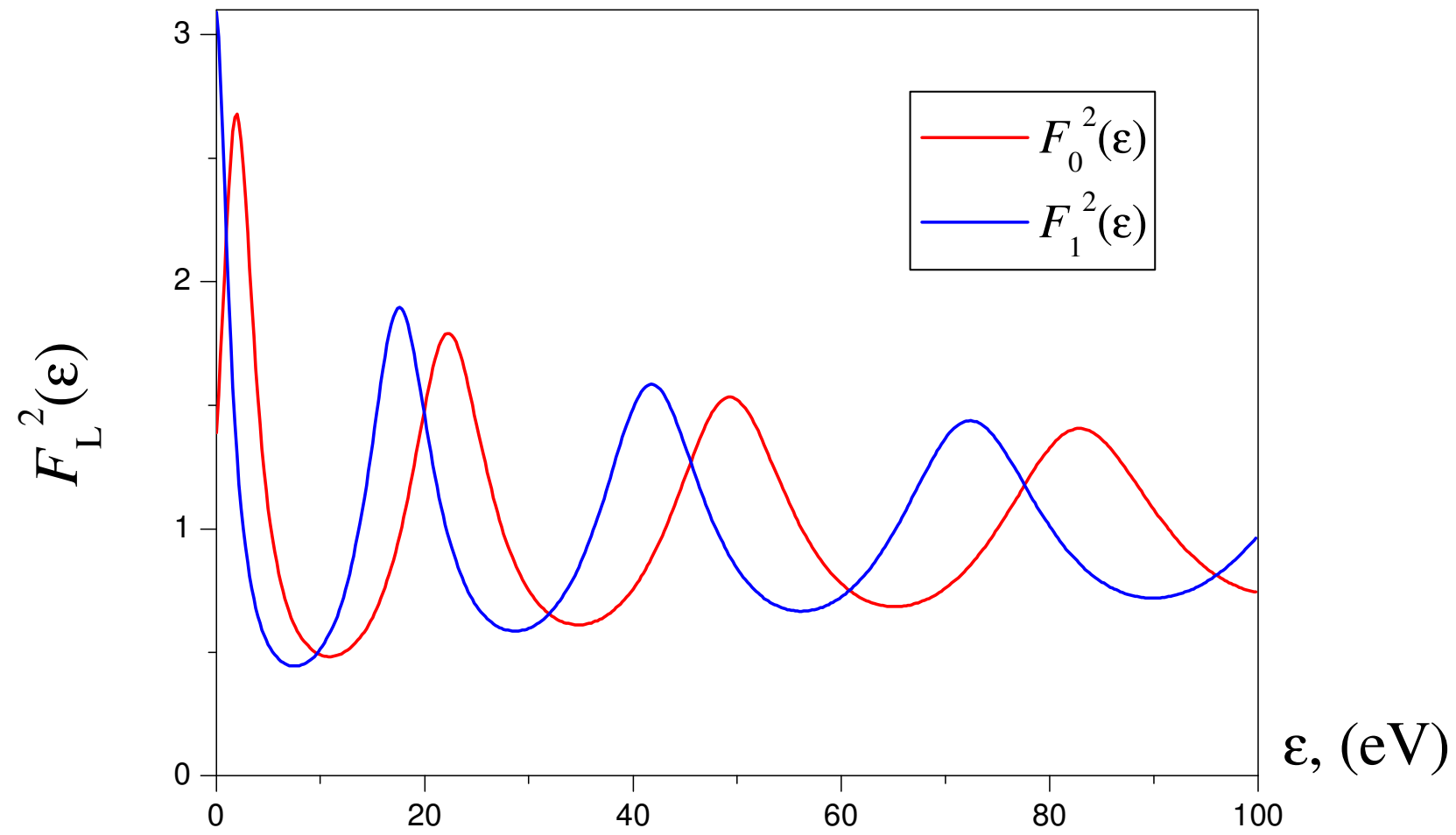

Fig. 17. Reflection factors as functions of the photoelectron energy.

The next figure presents combined action of two factors, $|F_o(k)|^2|F_1(k')|^2$ for a number of $\omega$ values, from 80 to 180 eV. Without photoelectrons' reflection by the $C_{60}$ shell, the differential in $\varepsilon$ two-electron spectrum is presented by an U-shape curve. The reflections strongly modulate the curve, as is seen from Fig. 18.

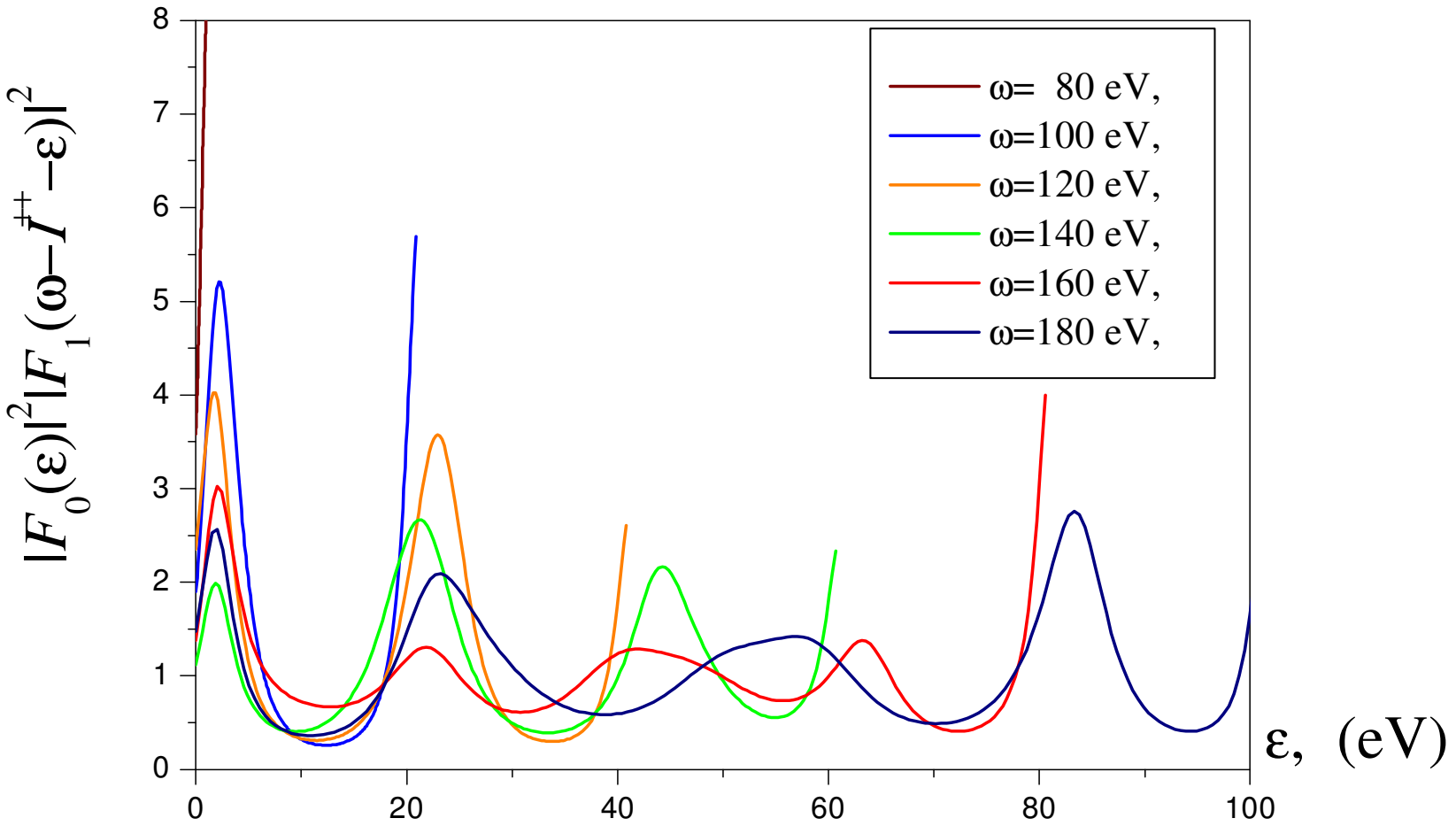

Fig.18. Modification of the two-electron spectrum under the action of the $C_{60}$ shell.

Of interest is the ratio $R_c(\omega)$ of total two-electron photoionization cross-sections with and without effect of the $C_{60}$ shell, $\sigma^{++}(\omega)$ and $\sigma^{a++}(\omega)$, that according to (22) is given by the following relation:

$$R_c(\omega) \equiv \int_0^{\omega-I^{++}} |F_0(\varepsilon)|^2 |F_1(\omega-I^{++}-\varepsilon)|^2 \, d\sigma^{a++}(\omega,\varepsilon) \Big/ \int_0^{\omega-I^{++}} d\sigma^{a++}(\omega,\varepsilon). \qquad (24)$$

Thus, to calculate this ratio, we need to know the two-electron differential in photoelectron energy cross-section. This is a separate problem. Most reliably $d\sigma^{a++}(\omega,\varepsilon)/d\varepsilon$ is theoretically calculated at high $\omega$ (see [26] and references therein).

To estimate $R_c(\omega)$, we took the results obtained in [26] and extrapolated it up to relatively small $\omega$, $\omega \geq I^{++}$. The ratio obtained is denoted as $R_c^{(h)}(\omega)$. At low $\omega$ the photoelectron energy distribution becomes almost $\varepsilon$-independent. As a result, one has for the low-energy ratio $R_c^{(l)}(\omega)$

$$R_c^{(l)}(\omega) \equiv \int_0^{\omega-I^{++}} |F_0(\varepsilon)|^2 |F_1(\omega - I^{++} - \varepsilon)|^2 \, d\varepsilon / (\omega - I^{++}). \tag{25}$$

Fig. 19 presents results of calculations for $R_c^{(h)}(\omega)$ and $R_c^{(l)}(\omega)$. It is remarkable that the corresponding functions starting from 110 eV become almost indistinguishable. This permits to assume that the effect of $C_{60}$ shell upon $R_c(\omega)$ is at any $\omega$ well reproduced by either $R_c^{(h)}(\omega)$ or $R_c^{(l)}(\omega)$.

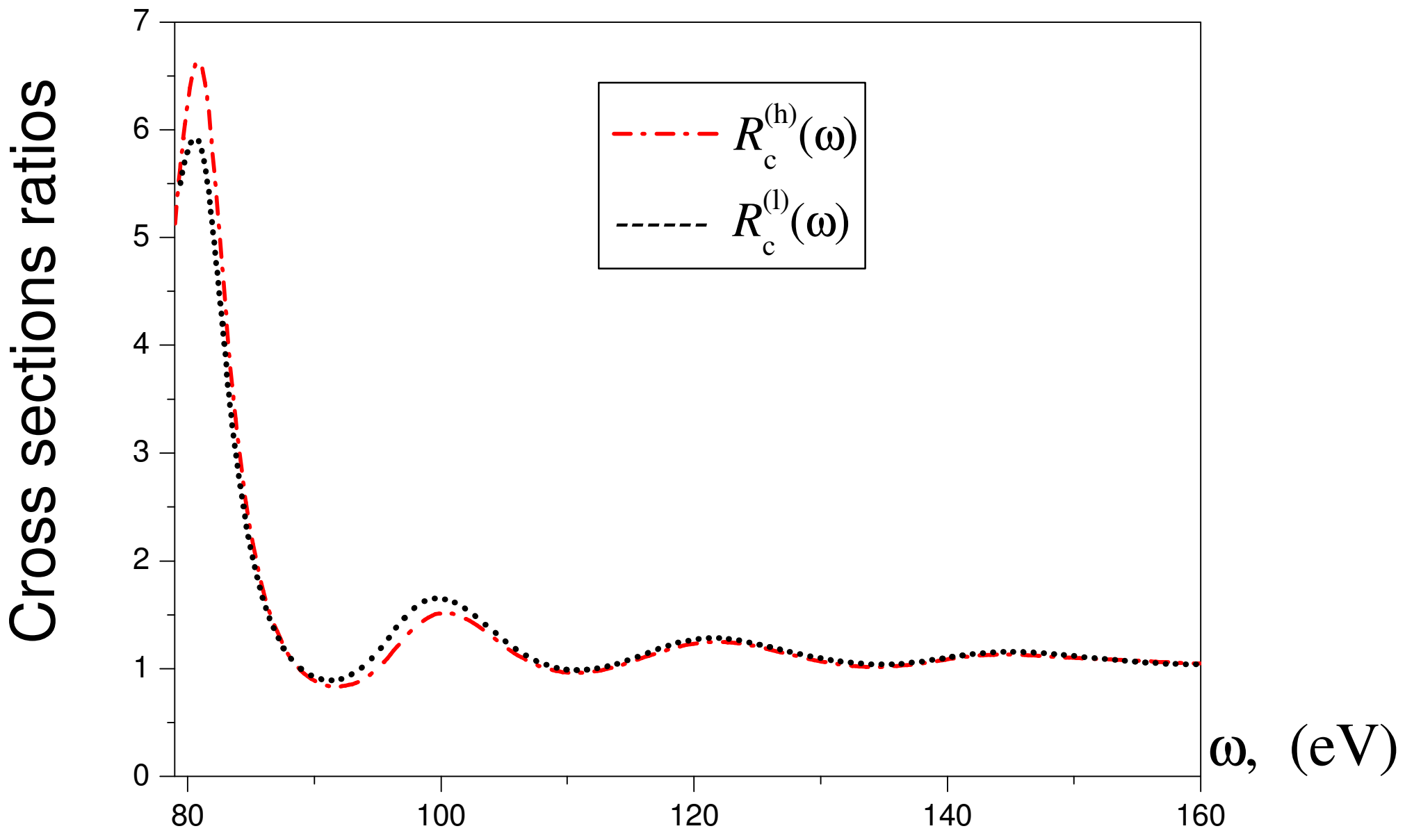

Fig. 19. Ratio endohedral-to-isolated He atom two-electron photoionization cross-section

**10. Conclusions and Perspectives**

We have investigated here effects of fullerenes shell upon photoionization of and vacancy decay in endohedral atoms. We demonstrate prominent modifications of the total and differential in angle and energy photoionization cross-sections that come from reflection of photoelectron waves by static $C_{60}$ potential and modification of the incoming electromagnetic field due to $C_{60}$ electron shell polarization. We found strong modifications of the vacancy decay probabilities and investigated the role of opening of new channels. The results were expressed via experimentally observable characteristics of the fullerene: its electron affinity and photoabsorption cross-section, along with the assumption that the fullerenes radius is much bigger than that of the atom and the fullerenes shell is thin as compared to its radius.

Modification of the incoming photon beam field and vacancy decay probabilities are mainly due to virtual or real excitation of the $C_{60}$ dipole Giant resonance. Concrete results were obtained for $C_{60}$, but qualitatively they are valid for other fullerenes, such as e.g. $C_{70}$ or $C_{76}$. One has to have in mined that many of the fullerenes are essentially deformed.

Such characteristics of endohedrals' photoionization as non-dipole anisotropy parameters are not properly considered in this paper since we do not have experimental data or reliable theoretical calculations of the quadrupole excitation spectrum. This is one of the problems that is of interest to study using the zero-thickness model. Another task is to introduce resistance of the fullerenes shell against alteration of its radius and to take into account the radius variation when an atom is stuffed inside a fullerene.

We need also to develop an approach that would permit on the same level of simplicity to describe inelastic scattering of photoelectrons with the fullerene shell and the shell's electron transitions into vacancies in endohedral atoms.

As soon as we will be able to treat the non-dipole excitations of the fullerenes shell and its resistance toward stretching, we could investigate the fast electron – endohedral atoms inelastic scattering and its characteristics – Generalized oscillator strengths. If good and simple approximation for wave functions of collectivized electrons in the frame of the "orange skin" model would be developed, it becomes possible to treat accurately enough those endohedral atoms vacancy decay, in which the vacancy in an endohedral is occupied by fullerenes electron. It would be possible to treat fullerenes and endohedrals multi-electron ionization with emission of fullerenes electrons.

One has to have in mind that ideal spherical symmetry is a essentially simplifying model: even for $C_{60}$ the octahedral structure adds some extra terms to the angular anisotropy of photoelectrons. It is essential that due to high symmetry of $C_{60}$ corrections of relatively high order in angular momentum of the field acting upon photoelectrons become important.

## 11. Acknowledgement

I am grateful to my co-workers Professors A. S. Baltenkov and V. B. Mandelzweig and Dr. E. Liverts, together with whom most results presented above, were obtained. I acknowledge the financial assistance of the Israeli Science Foundation, grant 174/03 and the Hebrew University Intramural Funds.